\documentclass[
twocolumn, 
twocolappendix,
numberedappendix,tighten,
]{aastex631mod}
\usepackage{multirow}
\pdfoutput=1 \usepackage{amsmath}
\usepackage{amstext}
\usepackage[T1]{fontenc}
\usepackage{apjfonts} 
\usepackage{booktabs}
\usepackage{listings}
\usepackage{colortbl}
\usepackage{xcolor}
\usepackage[encapsulated]{CJK}
\usepackage{xspace}
\usepackage[all]{hypcap}

\usepackage{hyperref}
\usepackage{nameref}

\providecommand{\sorthelp}[1]{}

\newcommand{\cntext}[1]{\begin{CJK}{UTF8}{gbsn}#1\end{CJK}}

\newcommand{\polspice}{\texttt{PolSpice}\xspace}
\newcommand{\sroll}{\texttt{SRoll2}\xspace}
\newcommand{\ghz}{\,GHz\xspace}

\defcitealias{Li23}{L23}
\defcitealias{pagano19}{P20}

\graphicspath{{./}{figures/}}
\definecolor{citecolor}{rgb}{0.08,0.30,0.85}
\hypersetup{citecolor=citecolor,filecolor=cyan}
\newcommand{\inprep}[1]{{\color{citecolor} #1 in prep.}}

\shorttitle{CLASS 90\,GHz Results}
\shortauthors{Li, Eimer et al.}

\begin{document}
\title{A Measurement of the Largest-Scale CMB $E$-mode Polarization with CLASS}
\newcommand{\jhu}{The William H. Miller III Department of Physics and Astronomy, Johns Hopkins University, 3701 San Martin Drive, Baltimore, MD 21218, USA}
\newcommand{\ucsc}{Departamento de Ingenier\'{i}a El\'{e}ctrica, Universidad Cat\'{o}lica de la Sant\'{i}sima Concepci\'{o}n, Alonso de Ribera 2850, Concepci\'{o}n, Chile}
\newcommand{\villanova}{Department of Physics, Villanova University, 800 Lancaster Avenue, Villanova, PA 19085, USA}
\newcommand{\goddard}{NASA Goddard Space Flight Center, 8800 Greenbelt Road, Greenbelt, MD 20771, USA}
\newcommand{\uchicago}{Department of Astronomy and Astrophysics, University of Chicago, 5640 South Ellis Avenue, Chicago, IL 60637, USA}
\newcommand{\kicp}{Kavli Institute for Cosmological Physics, University of Chicago, 5640 South Ellis Avenue, Chicago, IL 60637, USA}
\newcommand{\puci}{Instituto de Astrof\'isica, Facultad de F\'isica, Pontificia Universidad Cat\'olica de Chile, Avenida Vicu\~na Mackenna 4860, 7820436, Chile}
\newcommand{\pucc}{Centro de Astro-Ingenier\'ia, Facultad de F\'isica, Pontificia Universidad Cat\'olica de Chile, Avenida Vicu\~na Mackenna 4860, 7820436, Chile}
\newcommand{\argonne}{High Energy Physics Division, Argonne National Laboratory, 9700 S. Cass Avenue, Lemont, IL 60439, USA}
\newcommand{\upenn}{Department of Physics and Astronomy, University of Pennsylvania, 209 South 33rd Street, Philadelphia, PA 19104, USA}
\newcommand{\ucboulder}{Department of Astrophysical and Planetary Sciences, University of Colorado, 2000 Colorado Avenue, Boulder, CO 80309, USA}
\newcommand{\cfa}{Center for Astrophysics, Harvard \& Smithsonian, 60 Garden Street, Cambridge, MA 02138, USA}
\newcommand{\oslo}{Institute of Theoretical Astrophysics, University of Oslo, P.O. Box 1029 Blindern, N-0315 Oslo, Norway}
\newcommand{\MIT}{MIT Kavli Institute, Massachusetts Institute of Technology, 77 Massachusetts Avenue, Cambridge, MA 02139, USA}
\newcommand{\cepia}{CePIA, Astronomy Department, Universidad de Concepción, Casilla 160-C, Concepción, Chile}
\newcommand{\umbc}{University of Maryland, Baltimore County, 1000 Hilltop Circle, Baltimore, MD 21250, USA}
\newcommand{\lanl}{Space Remote Sensing and Data Science, Los Alamos National Laboratory, Los Alamos, NM 87545, USA}
\newcommand{\nist}{Quantum Sensors Division, NIST, 325 Broadway, Boulder, CO 80305, USA}

\author[0000-0002-4820-1122]{Yunyang Li (\cntext{李云炀}\!\!)}
\affiliation{\kicp}
\affiliation{\jhu}
\correspondingauthor{Yunyang Li}
\email{yunyangl@uchicago.edu}

\author[0000-0001-6976-180X]{Joseph~R. Eimer} \affiliation{\jhu}
\author[0000-0002-8412-630X]{John~W. Appel} \affiliation{\jhu}
\author[0000-0001-8839-7206]{Charles~L. Bennett} \affiliation{\jhu}
\author{Michael~K. Brewer} \affiliation{\jhu}
\author[0000-0003-2682-7498]{Sarah~Marie Bruno} \affiliation{\jhu}
\author[0000-0001-8468-9391]{Ricardo Bustos}\affiliation{\ucsc}
\author[0000-0001-8144-556X]{Carol Yan Yan Chan}\affiliation{\jhu}
\author[0000-0003-0016-0533]{David~T. Chuss}\affiliation{\villanova}
\author[0000-0002-7271-0525]{Joseph~Cleary}\affiliation{\jhu}
\author[0000-0002-1708-5464]{Sumit Dahal}\affiliation{\goddard}\affiliation{\jhu}
\author[0000-0003-3853-8757]{Rahul Datta}\affiliation{\uchicago}
\author[0000-0002-0552-3754]{Jullianna Denes~Couto}\affiliation{\jhu}
\author[0000-0002-3592-5703]{Kevin~L. Denis}\affiliation{\goddard}
\author{Rolando D\"unner}\affiliation{\puci}\affiliation{\pucc}
\author[0000-0002-4782-3851]{Thomas Essinger-Hileman}\affiliation{\goddard}
\author[0000-0003-1248-9563]{Kathleen Harrington}\affiliation{\argonne}\affiliation{\uchicago}
\author[0000-0001-9238-4918]{Kyle Helson}\affiliation{\umbc}\affiliation{\goddard}
\author[0000-0002-2781-9302]{Johannes Hubmayr}\affiliation{\nist}
\author[0000-0001-7466-0317]{Jeffrey Iuliano}\affiliation{\jhu}
\author{John Karakla}\affiliation{\jhu}
\author[0000-0003-4496-6520]{Tobias~A. Marriage}\affiliation{\jhu}
\author[0000-0002-2245-1027]{Nathan~J. Miller}\affiliation{\goddard}\affiliation{\jhu}
\author[0000-0002-1371-5334]{Carolina Morales~Perez}\affiliation{\jhu}
\author[0000-0002-8224-859X]{Lucas~P. Parker}\affiliation{\lanl}
\author[0000-0002-4436-4215]{Matthew~A. Petroff}\affiliation{\cfa}
\author[0000-0001-5704-271X]{Rodrigo A. Reeves}\affiliation{\cepia}
\author[0000-0003-4189-0700]{Karwan Rostem}\affiliation{\goddard}
\author[0009-0001-1748-7877]{Caleigh Ryan}\affiliation{\jhu}
\author[0000-0001-7458-6946]{Rui Shi (\cntext{时瑞}\!\!)}\affiliation{\jhu}
\author[0000-0002-2798-2943]{Koji Shukawa}\affiliation{\jhu}
\author[0000-0003-3487-2811]{Deniz A. N. Valle}\affiliation{\jhu}
\author[0000-0002-5437-6121]{Duncan~J. Watts}\affiliation{\oslo}
\author[0000-0003-3017-3474]{Janet~L. Weiland}\affiliation{\jhu}
\author[0000-0002-7567-4451]{Edward~J. Wollack}\affiliation{\goddard}
\author[0000-0001-5112-2567]{Zhilei Xu (\cntext{徐智磊}\!\!)}\affiliation{\MIT}
\author[0000-0001-6924-9072]{Lingzhen Zeng}\affiliation{\cfa}

\received{2025 January 20}
\revised{2025 March 25}
\accepted{2025 March 28}
\submitjournal{\apj}
\begin{abstract}
We present measurements of large-scale cosmic microwave background (CMB) $E$-mode polarization from the Cosmology Large Angular Scale Surveyor (CLASS) 90\ghz data. 
Using 115 det-yr of observations collected through 2024 with a variable-delay polarization modulator, we achieved a polarization sensitivity of $82\,\mathrm{\mu K\,arcmin}$, comparable to Planck at similar frequencies (100 and 143\ghz). 
The analysis demonstrates effective mitigation of systematic errors and addresses challenges to large-angular-scale power recovery posed by time-domain filtering in maximum-likelihood map-making. 
A novel implementation of the pixel-space transfer matrix is introduced, which enables efficient filtering simulations and bias correction in the power spectrum using the quadratic cross-spectrum estimator. 
Overall, we achieved an unbiased time-domain filtering correction to recover the largest angular scale polarization, with the only power deficit, arising from map-making non-linearity, being characterized as less than 3\%.
Through cross-correlation with Planck, we detected the cosmic reionization at $99.4\%$ significance and measured the reionization optical depth $\tau=0.053 ^{+0.018}_{-0.019}$, marking the first ground-based attempt at such a measurement. 
At intermediate angular scales ($\ell > 30$), our results, both independently and in cross-correlation with Planck, remain fully consistent with Planck's measurements.
\end{abstract}

\keywords{ 
    \href{http://astrothesaurus.org/uat/322}{Cosmic microwave background radiation (322)};  
    \href{http://astrothesaurus.org/uat/1146}{Observational Cosmology (1146)}; 
    \href{http://astrothesaurus.org/uat/1277}{Polarimeters (1277)};
    \href{http://astrothesaurus.org/uat/1383}{Reionization (1383)};
    }

\section{Introduction}
The largest angular scale modes of the cosmic microwave background (CMB) polarization contain critical cosmological information that, to date, has only been observable through space-based missions \citep{bennett13,planck2016-l01}.
Specifically, the epoch of reionization enhances the polarization signal on large angular scales, encoding most of its information in the dominant $E$-mode power and boosting the detectability of inflationary gravitational waves in the $B$ modes.
While the search for primordial $B$ modes could also be attempted from the ground at intermediate angular scales \citep[$\ell\gtrsim 30$;][]{abs18,pb20bmode,BK21}, direct determination of the global properties of the reionization history from CMB $E$-mode polarization is only possible from $\ell\lesssim 20$ \citep{zaldarriaga97}.

The reionization period affects CMB radiation through scattering by free electrons. 
On the largest scales, the $E$-mode polarization is enhanced by a factor approximately proportional to $\tau^2$ \citep{zaldarriaga97} where $\tau$ is the reionization optical depth. 
At smaller scales, this scattering suppresses the anisotropy spectrum by a factor of $e^{-2\tau}$ and would also leave detectable imprints in the temperature field if reionization happened inhomogeneously \citep{gruzinov98,dvorkin09,namikawa21,raghunathan24}.
To the leading order, a measurement of $\tau$ can constrain the reionization epoch given assumptions of the phase transition \citep{Shull2008,tanh}. 
More details of the reionization process can be reconstructed from the full shape of large-scale $E$-mode spectrum \citep{mortonson08,watts20}, which can be further complemented by neutral hydrogen probes and direct observations of the ionizing sources to paint a complete picture of the reionization history \citep{munoz24,paoletti24}.
Because reionization also bolsters the $B$-mode spectrum, accurately modeling reionization through $E$-mode measurements is crucial for future characterizations of inflationary $B$ modes on the largest scales 
 \citep{mortonson08-sr, jiang&namikawa24}.

The first measurement of $\tau$ from CMB data was obtained by WMAP using the $TE$ correlation \citep{kogut03} and later refined with the inclusion of $EE$ power spectra, yielding $\tau=0.089\pm0.014$ \citep{hinshaw13}. 
The current strongest constraint on $\tau$ comes from the Planck High-Frequency Instrument (HFI) polarization data. Across various map-making and parameter inference approaches, the $\tau$ value inferred from Planck HFI low-$\ell$ $EE$ spectra remains consistent with approximately $0.058$ at the $10\%$ level (\citealt{pagano19}, hereafter, \citetalias{pagano19}; \citealt{tristram24}).
Data combinations of WMAP and/or Planck Low-Frequency Instrument (LFI) while leaving out HFI (except for dust cleaning) typically find intermediate $\tau$ values between $0.06-0.07$ \citep{lattanzi17,weiland18,planck2016-l05,natale20,paradiso23}.

In addition to its astrophysical implications, the precise value of $\tau$ is crucial in breaking degeneracies with the scalar fluctuation amplitude $A_\mathrm{s}$ through $A_\mathrm{s}e^{-2\tau}$, to which the small-scale CMB anisotropy amplitude is most sensitive. 
This is particularly important for inference of parameters that depend on the absolute amplitude of $A_\mathrm{s}$ and help elucidate tensions in cosmic concordance.
A smaller value of $\tau$ would imply a lower $A_\mathrm{s}$ thereby alleviating the $\sigma_8$ tension between the CMB anisotropy measurement and that from large-scale structure probes at lower redshifts \citep{planck2014-a10}. 
A larger $\tau$ value might be related to the lensing anomaly that requires a higher lensing power (thus $A_\mathrm{s}$) to explain the excessive peak smearing in Planck temperature spectrum \citep[e.g.,][]{planck2016-l06,rosenberg22,addison24}. 
Indeed, $\tau$ determined from high-$\ell$ spectra and CMB lensing, completely independent of polarization data, shows an upshift toward $0.08$ at $20\%$ precision \citep{planck2016-l06,giare24} whereas marginalizing over a phenomenological $A_L$ parameter \citep{calabrese08} to counteract the lensing excess restores agreement with low-$\ell$ polarization measurements \citep{planck2016-LI,couchot17-AL}.
The $\tau-A_\mathrm{s}$ degeneracy also plays a role in the inference of the sum of neutrino masses through its effect in suppressing structure growth \citep{Bond1980}. 
Improved precision on $\tau$ would thus strengthen neutrino mass constraints \citep{allison15neutrinos,loverde24} and inform recent hints of neutrino mass trending below the bounds established by oscillation experiments \citep{desi24-FS,green&meyer24}.

In the interim between Planck and future space-based missions \citep{pico19,litebird22}, various efforts have been undertaken or proposed to improve low-$\ell$ polarization measurements, both from the ground \citep{lspe20,groundbird20} and from balloon platforms \citep{lspe20,errard22,may24}.
The Cosmology Large Angular Scale Surveyor \citep[CLASS;][]{essinger-hileman14spie} is one of the ground-based efforts located in the Atacama Desert in Chile, where it has been observing since 2016.
CLASS is an array of telescopes operating at frequencies around 40, 90, 150, and 220\ghz.
Its ability to observe large angular-scale variations in polarized sky emission is enabled by rapid front-end modulation using a variable-delay polarization modulator \citep[VPM;][]{chus12vpm,harrington18spie,harrington21},\footnote{The VPM was used for all data in this analysis. CLASS recently started using a reflective half-wave plate (see Section~\ref{ssec:survey-timeline}) to test alternative polarization modulation techniques and to increase the linear polarization sensitivity.} which also suppresses spurious instrumental polarization and provides unique sensitivity to circular polarization \citep{petroff20,padilla20,eimer23}.
The capability of CLASS to make scientific measurements on the largest angular scales has recently been demonstrated using the 40\ghz data \citep{padilla20,eimer23,shi24}.
In this paper, we describe the processing of a subset of CLASS 90\ghz linear polarization data from three years of observation taken through 2024, showcase the analysis techniques required for large-angular-scale recovery, and present the first ground-based measurement of the reionization optical depth through cross-correlation with Planck.

Section~\ref{sec:data} begins by presenting the CLASS observation strategy and the 90\ghz data used in the analysis, followed by a description of the map-making process in Section~\ref{sec:map-making}. The internal validation and external cross-checks are described in Section~\ref{sec:validation}, and the measurement of the reionization optical depth is detailed in Section~\ref{sec:tau}, with conclusions drawn in Section~\ref{sec:conclusion}.
Throughout this work, the best-fit $\Lambda$CDM cosmology from the Planck 2018 \texttt{TTTEEE+lowE} data combination \citep{planck2016-l06} is taken as the fiducial model, specifically $10^9A_\mathrm{s} e^{-2\tau}=1.884\pm0.012$ and $\tau=0.0544$.

\section{CLASS and the 90 GHz Data Set}\label{sec:data}
\begin{figure*}
    \includegraphics[width=\linewidth, ]{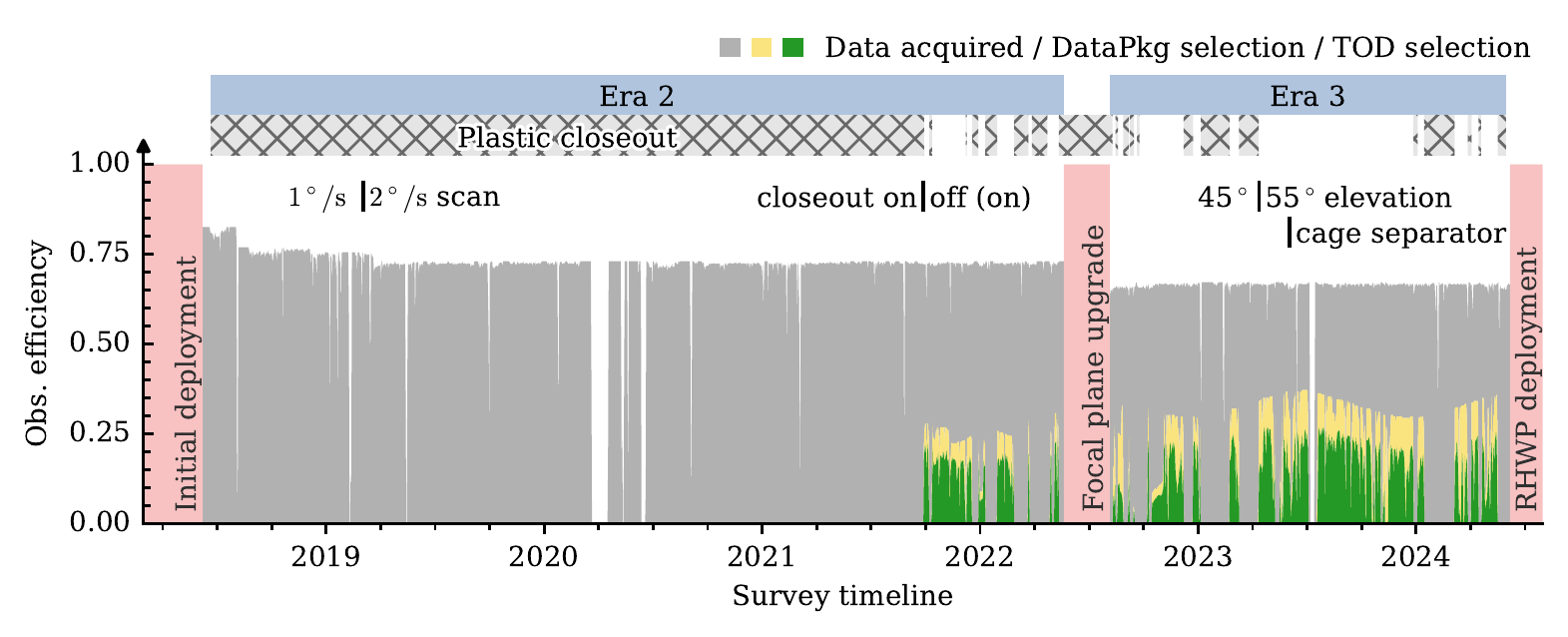}
    \caption{
    CLASS daily observation efficiencies (see definition in text) of the 90\ghz telescope since the initial deployment in 2018 through the recent modulator replacement in June 2024. 
    The gray region represents the total detector uptime when the modulator (VPM) was also operational. The yellow region indicates the data volume initially selected at the DataPkg level while the green region shows the fraction retained after the TOD-level selection (see Section~\ref{ssec:data-selection}).
    The telescope “Era” shown in the blue band at the top denotes major changes in the instrument configuration. 
    Gray hatched regions correspond to periods when a plastic closeout was installed.
    The closeout was first removed on 2021-09-27 but was occasionally reinstalled to protect the instrument from adverse weather conditions. 
    The data with the closeout installed are conservatively left out for the current analysis. 
    Critical changes in the instrument configuration are denoted by vertical bars, with annotations positioned in the direction of their corresponding descriptions.
}
    \label{fig:class-overview}
\end{figure*}
\subsection{CLASS Telescope}
All four CLASS telescopes share a similar design. Two telescopes (e.g., 40\ghz and 90\ghz) are paired on the same tri-axial mount, enabling continuous $720^\circ$ azimuth scanning and daily boresight rotations from $-45^\circ$ to $45^\circ$ in $15^\circ$ increments. 
Each telescope employs a reflective polarization modulator as its first optical element \citep[a VPM for the data discussed here;][]{harrington18spie}, followed by two ambient-temperature ellipsoidal mirrors that direct radiation into the cryogenic receiver \citep{eimer12spie}. 
The signal then passes through IR-blocking filters in multiple cooling stages \citep{iuliano2018spie} and is focused by dielectric lenses onto the focal plane. The focal plane uses smooth-walled feedhorns \citep{zeng10} to couple the signal to orthogonal probe antennas, each direction ultimately terminating in a transition-edge-sensor (TES) bolometer cooled to approximately 50\,mK \citep{chuss12tes, rostem16spie}.
The 90\ghz telescope initially fielded seven modules each with 37 dual-polarization detectors \citep[74 TES bolometers;][]{dahal18spie,datta23,nunez24}.
We refer the readers to the references cited herein for more instrumentation details.

\subsection{Survey Timeline}\label{ssec:survey-timeline}
The first of the two 90\ghz telescope arrays was deployed in June 2018 and has been observing since then.
This telescope was installed on the same mount as the 40\ghz telescope, which began operation in 2016 \citep[][hereafter, \citetalias{Li23}]{Li23}. 
Therefore, following the conventions in previous CLASS literature, we denote the first 90\ghz observing season since 2018 as ``Era 2''.
In response to stability issues in the TES bias range and low optical efficiency \citep{dahal22}, the four (out of seven) detector modules with the worst performance were replaced in August 2022. These newly designed modules incorporated updated electromagnetic and electro-thermal circuit modifications, resulting in an overall detector optical efficiency exceeding 90\% \citep{nunez24}---the observation period since the focal plane upgrade is denoted as ``Era 3''.
Era-3 operations concluded in June 2024 to prepare for the installation of a reflective half-wave plate (RHWP) for the 90\ghz telescope \citep{eimer2022spie,shi24-spie}.
This paper focuses only on the 90\ghz data from the first 90\ghz telescope in Era 2 and Era 3, which used the VPM as the modulator.

The timeline and observation efficiency (defined as the ratio between the achieved integration time and a 24-hour observation with a fully populated focal plane) of the 90\ghz observation is summarized in Figure~\ref{fig:class-overview}, with important events and configuration changes annotated and discussed below:

\paragraph{Azimuth Scan Speed Increase} CLASS conducts constant-elevation azimuth scan during CMB observation for the entire 90\ghz operation. 
The scanning speed was increased from $1$ to $2\,\mathrm{deg\,s^{-1}}$ on 2019-03-04, which was found to improve the low-frequency stability of the data (\inprep{Cleary et al.}).

\paragraph{Closeout Removal} A thin plastic environmental seal, known as the “closeout,” was initially installed at the base of the telescope’s forebaffle extension during routine CMB observations. However, deformation of this plastic closeout under wind load induces a spurious polarization signal that is correlated among all detectors and appears quasi-synchronous in azimuth due to the prevalent wind direction \citepalias[see][for details on its impact at 40\ghz]{Li23}. Since 2021-09-27, the closeout was removed for CMB observations except under adverse weather conditions. The periods during which the closeout was installed are indicated by hatched boxes in Figure~\ref{fig:class-overview}.

\begin{figure}[t!]
    \centering
    \includegraphics[width=\linewidth]{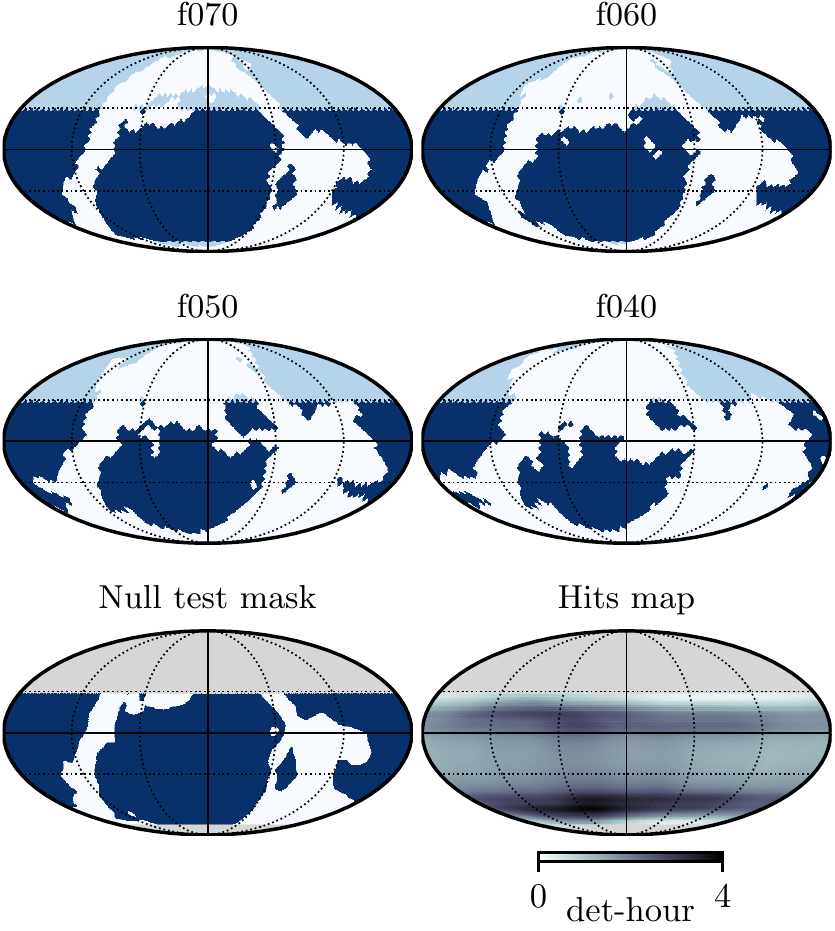}
        \caption{Galactic avoidance masks used for this analysis (first 5 panels) and the 90\ghz hits map (last panel) in celestial coordinates.
        The first two rows show the low-resolution Galactic avoidance masks 
        at $N_\mathrm{side}=16$, with retained sky fraction ranging from 70\% to 40\%. 
        The region not surveyed by CLASS but used for Planck analysis is painted in lighter blue.
        We use the ``\texttt{f0XX}'' notation to denote the retained sky fraction after Galactic avoidance for both CLASS and Planck masks, with the understanding that the corresponding masks for CLASS data cover smaller sky fractions due to the declination limits.
        The bottom-left panel shows the mask used for the null tests, which is the same as \texttt{f070} but constructed at the native $N_\mathrm{side}=256$ resolution. 
        The polarization hits map in the last panel shows the total integration time (not accounting for the linear polarization modulation efficiency) on each quarter-degree $N_\mathrm{side}=256$ pixel.
        \label{fig:masks}}
\end{figure}

\paragraph{Azimuth Scan Elevation} The scan elevation was increased from $45^\circ$ to $55^\circ$ on 2023-04-12. 
This change was designed to 1) examine the level of contribution from ground pickup to the scan-correlated signals, and 2) to improve the sky coverage in the mid-declination range (see Figure~\ref{fig:masks}). 
Preliminary analysis showed little difference in the signal stability due to this elevation change. This suggests that ground emission is not a dominant source of the observed scan-correlated linear polarization signals measured by the CLASS 90\ghz channel (\inprep{Chan et al.}). 
\paragraph{Cage Separator}
Analyses of the 40\ghz and 90\ghz Sun-centered maps revealed a fan-shaped sidelobe at $-70\,\mathrm{dB}$ pointing in the direction of the co-mounted telescope \citepalias[][Figure 5]{Li23}. 
This was attributed to the opening in the optical path between the two telescopes. On \mbox{2023-06-02}, these sidelobes were removed by installing an absorbing wall that separates the two telescopes within the comoving ground screen.

\subsection{Data Selection}\label{ssec:data-selection}
The data selection follows closely the treatment for the 40\ghz data in two steps \citepalias{Li23}.

In the first step, 10-minute data segments, each called a \emph{DataPkg} \citep{petroff20spie}, were selected based on observation conditions and instrument status. 
In contrast to the 40\ghz selection criteria, we did not include any daytime data and restricted the Sun position to be below the horizon. 
We also discarded data acquired while the plastic closeout was installed.\footnote{Although this systematic can be removed through filtering for the 40\ghz data \citepalias{Li23}, we encountered additional difficulties in modeling the low-$\ell$ noise under our current mitigation strategy, yielding only marginal improvements to low-$\ell$ sensitivity when including this data set. 
Nevertheless, it remains unclear whether this limitation stems solely from the closeout since we observed low-$\ell$ performance improvement over time that did not align exactly with the closeout status.} 
This resulted in nearly half of the survey data being removed, including all data prior to 2021-09 along with a substantial fraction collected thereafter.
The discarded data set could still be used to improve mid-$\ell$ sensitivity and may be recoverable for future low-$\ell$ studies with revised analysis methods.
The selected DataPkgs at this stage, amounting to 190 det-yr, are shown as the yellow shaded region in Figure~\ref{fig:class-overview}.

In the second step, DataPkgs from every observing night were concatenated to form a ``span'', which is the data unit for time-ordered data (TOD)-level operations including data cuts, calibration, demodulation, and filtering. 
A span typically comprises $\sim80$ DataPkgs for nighttime-only data.
Similar to the approach in \citetalias{Li23}, the TOD-level data selection was based on criteria including Moon position, precipitable water vapor magnitude, detector readout status, cryostat health, abnormal detector responses (as probed by the response to the VPM emission signal), and proximity to the azimuth scan turnarounds. 
In addition to the 40\ghz treatment, we implemented data cuts after demodulation. 
These cuts target variability or abnormal trends in the \textit{demodulated} data to improve specifically linear polarization measurement quality \citep{Li-thesis}. 
The final retained data used for map-making, totaling 115\,det-yr, are shown as the green shaded region in Figure~\ref{fig:class-overview}.

\subsection{Masks}
In this section, we summarize various Galactic masks that are used for this analysis.
The construction of the mask is the same as that used for the \sroll analysis \citepalias{pagano19} where the WMAP $K$-band and Planck 353\ghz polarization maps were smoothed to $7.5^\circ$ FWHM and scaled to 100 and 143\ghz (assuming synchrotron and thermal dust spectrum) respectively. 
The sum of the two polarization intensities was thresholded at different levels to create masks with different retained sky fractions. 
These masks are made at both \texttt{Healpix} \citep{healpix} $N_\mathrm{side}=256$ ($0.23^\circ$) and $N_\mathrm{side}=16$  ($3.7^\circ$) resolutions for analysis working at different scales.
When CLASS data are involved, we restrict the mask to be within the CLASS declination boundaries, $-74^\circ\leq \delta \leq28^\circ$.
These masks are denoted by ``\texttt{f0XX}'' where the number represents the retained sky fraction in percentage for full-sky maps (e.g., Planck), with the understanding that the corresponding sky fractions for CLASS masks are smaller.
For \texttt{f070}--\texttt{f040}, the CLASS mask sky fractions are 52\%, 45\%, 37\%, and 28\%, respectively.
These masks are shown in Figure~\ref{fig:masks}.

\section{Map-making}\label{sec:map-making}

\subsection{Demodulation}\label{ssec:demod}
The sky polarization signals are encoded in the raw detector data through the VPM modulation.
A demodulation process (\citealt{harrington21}; \citetalias{Li23}) was applied to recover the linear and circular polarization from calibrated raw data \citep{appel22}. 
For this analysis, we focus only on the linear polarization data, and the circular polarization data will be presented in future work (\inprep{Essinger-Hileman et al.})
The demodulation requires knowledge of both the instrument parameters (VPM model and detector bandpass) and the spectral index of the sky signal to separate the two polarization states cleanly.
While this was achieved through a brute-force optimization process for the 40\ghz result \citep{eimer23}, here we only used the fiducial model of the instrument.
Potential impacts of this ignorance include leakage of the atmospheric circular polarization that will be filtered from the demodulated data (Section~\ref{ssec:filtering}) and a bias in the linear polarization amplitude that will be absorbed by a final absolute calibration adjustment (Section~\ref{ssec:external-validation}).

The raw data sampling rate, approximately $200\,\mathrm{Hz}$, is designed to record the modulated signals (the 10\,Hz modulation frequency and its higher harmonics). The demodulated data were down-sampled to $2.5\,\mathrm{Hz}$ in this analysis to enable efficient downstream processing. 
However, this reduced rate is only marginally higher than the $0.62^\circ$ FWHM-beam-crossing frequency, causing the associated anti-aliasing filter to introduce non-negligible signal bias on scales near the pixel resolution. Simulations characterizing this effect are presented in Section~\ref{ssec:beam-window}, and the effect is incorporated into the total window function.

\subsection{Filtering}\label{ssec:filtering}

Like the 40\ghz data analysis, scan-correlated signals were detected in the demodulated linear polarization time stream that require filtering before map-making \citepalias{Li23}.
For the entirety of the 90\ghz data used for analysis, the telescope conducts two continuous rotations at a constant elevation, covering a total of $720^\circ$ in azimuth ($4\pi$, or a \textit{sweep}), before reversing direction for another sweep.
This scanning pattern creates two types of scan-correlated signals: those associated with the ground would have $2\pi$ periodicity and those associated with the scanning motion would have $8\pi$ periodicity.
Details of these spurious modes will be studied in a companion paper (\inprep{Chan et al.}).

Due to the scan reversal at turnarounds, the ground/sky\footnote{At short time separation, the sky signal is approximately azimuth-dependent.} signal would be wrapped around in the time domain and be only partially degenerate (the even-parity mode about the turnaround) with the azimuth-velocity-induced $8\pi$ signals.
This motivates the design of two sets of filter basis: 1) a set of filters to remove the first 12 harmonic modes of the $8\pi$ signals that contain azimuth-velocity-induced systematics, and 2) a set of filters in the azimuth domain to remove the first 12 harmonic modes of the $2\pi$ signals. 
The azimuth velocity filters are re-estimated every 8 sweeps, whereas the first and second 6 modes of the azimuth filters are fit over 30 sweeps and the entire span ($\sim 120$ sweeps), respectively.
In addition to the scan-correlated filters, we also included a set of low-order polynomial filters to improve the stability of the joint fit.
For each detector, the demodulated data ($\mathbf{d}$) were filtered as follows:
\begin{align}
    \mathbf{d} \rightarrow \mathbf{d} - \mathbf{V}(\mathbf{V}^T\mathbf{V})^{-1}\mathbf{V}^T\mathbf{d},
\end{align}
where $\mathbf{V}$ bundles together all filter components (including all models and all modes) as column vectors.
This is a linear operation whose impact on the signal bias will be characterized by linear operators at the map level (see Section~\ref{sssec:pix-tf}).

\subsection{Sky Maps}

\begin{figure*}[t!]
    \centering
    \includegraphics[width=\linewidth]{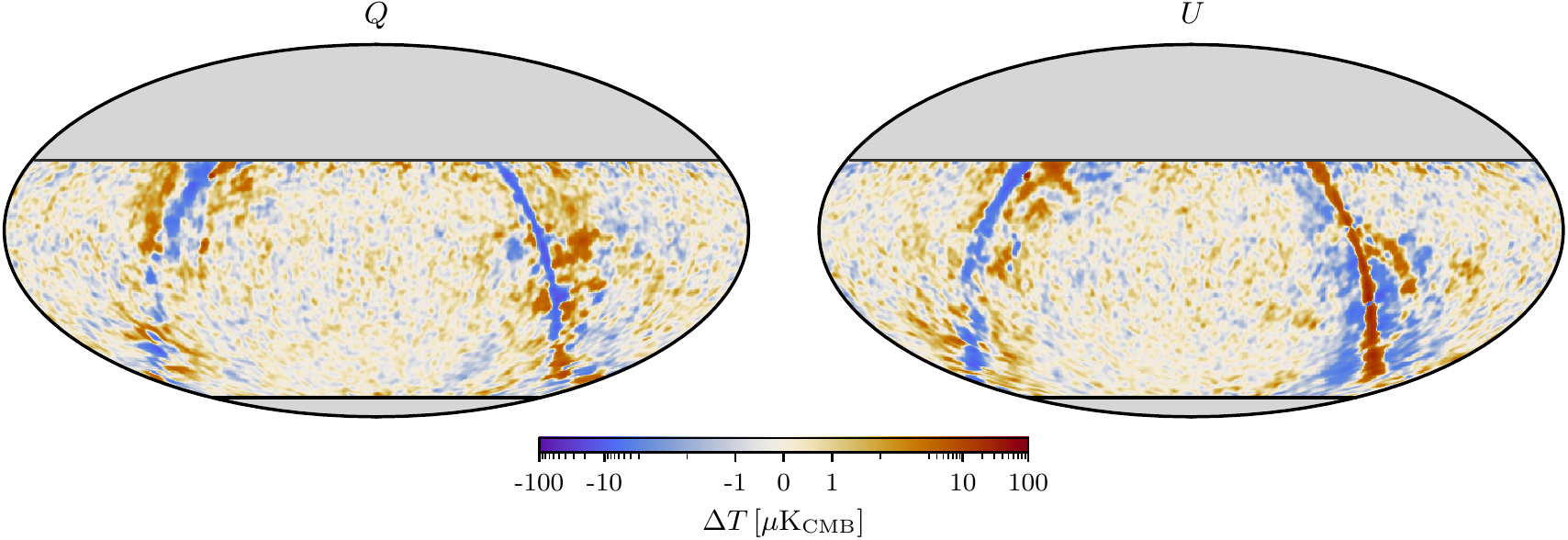}
        \caption{CLASS 90\ghz linear polarization maps shown in Celestial coordinates under Mollweide projection. 
        Gray regions indicate portions of the sky not surveyed by CLASS. These maps have been smoothed to 2$^\circ$ FWHM for visualization.
        \label{fig:maps}}
\end{figure*}
After filtering, the demodulated linear polarization data were weighted and projected onto the sky using a maximum-likelihood map-making algorithm (\citealt{dunner13,romero20}; \citetalias{Li23}).
The core of this method involves constructing noise covariance matrices for data segments that capture most of the temporal and focal plane correlations. 
For the 90\ghz data, a noise model was estimated approximately every two hours, balancing noise stability, computational cost, and model accuracy.
Given the compact form of the noise model, the map solution was obtained using the conjugate gradient method.
A challenge arises from the need to derive the noise model from the data themselves, which can introduce signal bias and disrupt linearity, especially when the \emph{time-ordered data} are less noise-dominated.
This was mitigated through an iterative template subtraction process, where the map solution from each iteration was subtracted from the time-ordered data before re-estimating the noise model for the next run (\citetalias{Li23}; \citealt{dunner13}). For the CLASS 90\ghz data, five template iterations were performed.
This method has been shown to \emph{largely} remove bias and restore linearity in map-making across nearly all scales for the 40\ghz data \citepalias{Li23}. 
A detailed characterization of its impact on the 90\ghz analysis is provided in Section~\ref{sssec:map-bias}.

The CLASS 90\ghz linear polarization maps are shown in Figure~\ref{fig:maps} in Celestial coordinates.
A map of the linear polarization integration time (the hits map) is shown in the last panel of Figure~\ref{fig:masks}.
The polarization map depth is $82\,\mathrm{\mu K\,arcmin}$, based on the angular power spectrum in the range $100<\ell<250$ where it is mostly flat.

\subsection{Transfer Functions}
\subsubsection{Beam Window Functions}\label{ssec:beam-window}
\begin{figure}
    \includegraphics[width=\linewidth]{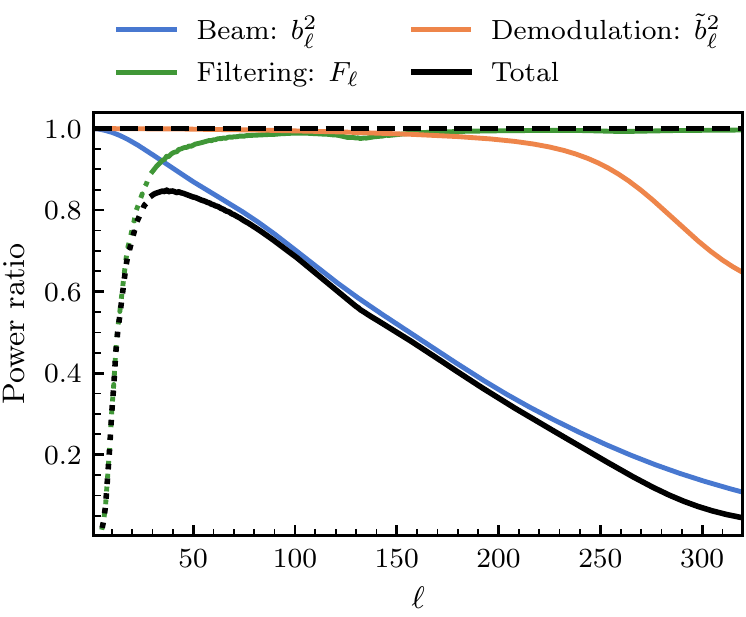}
    \caption{
    $EE$ power spectrum transfer functions. 
    The blue curve shows the beam window function ($b^2_\ell$), and the orange curve represents the effect of the anti-aliasing low-pass filtering before down-sampling the demodulated data ($\tilde{b}^2_\ell$).
    The green curve is the isotropic mapping transfer function ($F_\ell$) computed from the spectrum ratio between the filtered and unfiltered maps using \polspice.
    This calculation ignores mode coupling and polarization mixing. 
    Therefore, it is only valid for high-$\ell$ corrections, and care needs to be taken when interpreting $\ell \lesssim 30$ (represented by dotted lines).
    The black curve includes all effects above.
    The transfer function for the $B$ mode is almost identical.\label{fig:filtering_tf}
    }
\end{figure}

At high multipoles, the signal bias is largely influenced by the instrument beam and the ``digital beam” effect induced by the low-pass filter in the demodulation step (Section~\ref{ssec:demod}). 
The 90\ghz beam model, established from dedicated Jupiter observations, has an approximate FWHM of $0.62^\circ$ \citep{datta23}, and its corresponding beam window function, $b^2_\ell$, is shown as the blue curve in Figure~\ref{fig:filtering_tf}.

To characterize the digital beam from low-pass filtering, raw data simulations were demodulated, downsampled, and mapped following the same procedure as the real data \citepalias{Li23}. 
The resulting effect can be approximated as an isotropic window function in angular space, denoted as $\tilde{b}^2_\ell$, and is depicted as the orange curve in Figure~\ref{fig:filtering_tf}.

\subsubsection{Pixel-space Transfer Matrix}\label{sssec:pix-tf}
To accurately characterize the filtering effect at the map level and the power spectra level, we aim to build a transfer matrix to forward model the filtering in pixel space \citep{BKpure}.
A similar approach has been outlined for the CLASS 40\ghz data in \citetalias{Li23}, but here we refine the matrix estimation and provide more details regarding the assumptions made for this approximation.
Since the estimation is computationally prohibitive at the full resolution of the maps, and only the low-$\ell$ modes are of interest, we only constructed the matrix for maps at $N_\mathrm{side}=16$ (the pixel number scales as $N_\mathrm{side}^2$).
To convert the full-resolution maps into lower resolution, we introduce a downgrade operator $\mathcal{D}_{N_\mathrm{side}}$ to rebin maps onto $N_\mathrm{side}$ grid. 
The map-level operator \texttt{ud\_grade} from \texttt{Healpix} is deliberately avoided to prevent aliasing effects and polarization bias \citep{sullivan24}.
The downgrade operator $\mathcal{D}_{N_\mathrm{side}}$ first transforms the map into spherical harmonic space and then applies a cosine anti-aliasing filter \citep{benabed09,planck2014-a10}:

\begin{equation}
    f_\ell = 
    \begin{cases} 
    1&  \ell \leq N_\mathrm{side},\\
    \frac{1}{2}\Big(1+\sin \dfrac{\pi \ell}{2N_\mathrm{side}}\Big) &  N_\mathrm{side} < \ell \leq 3N_\mathrm{side}, \\
    0&  \ell > 3N_\mathrm{side}.
    \end{cases}\label{eq:fl}
\end{equation}
where $N_\mathrm{side}$ is the output resolution.
Finally, the $f_\ell$-applied spherical harmonics are converted back to the pixel space (at resolution $N_\mathrm{side}$) with the appropriate pixel window function applied.
When a series of downgrade operators are chained, it should be assumed that the low-pass filters and pixel window functions from previous steps have been corrected for.

Tracking only the sky signal in the map, the downgraded map can be expressed as
\begin{align}
    \tilde{\mathbf{m}}_{16} = \mathcal{D}_{16}\tilde{\mathbf{m}} &= \mathcal{D}_{16}\mathcal{M}(\mathbf{m}_{N_F} + \mathrm{sub.~pix.}) \label{eq:pix}\\
    &\approx \mathcal{D}_{16}\mathcal{R}\mathbf{m}_{N_F}\label{eq:tfpix-approx-2}\\
    & = \mathbf{F}_{\mathrm{16},N_F}\mathbf{m}_{N_F}\label{eq:tfpix-approx-3},
\end{align}
where $\tilde{\mathbf{m}}_{16}$ is the downgraded filtered map ($\tilde{\mathbf{m}}$) at $N_\mathrm{side}=16$, $\mathcal{M}$ is the map-making operator that includes projecting sky signals into the (demodulated) TOD, filtering, and map-making. 
The terms in the parentheses separate the signal into a component that is well sampled at $N_\mathrm{side}=N_F$, $\mathbf{m}_{N_F}$, and sub-pixel information that does not survive the downgrade operation.
$N_F$ is an adjustable parameter, but it must not be smaller than the output resolution  $N_\mathrm{side}=16$ in Equation~\ref{eq:pix}. 
A higher value provides a more accurate approximation, though at the expense of increased computational cost when constructing the filter matrix.
In line~\ref{eq:tfpix-approx-2}, we approximate Equation~\ref{eq:pix} by dropping the sub-pixel information and replacing the map-making operator with the reobservation operator $\mathcal{R}$.
The former is a good approximation since sub-pixel information should not be significantly aliased to larger scales in the map-making, i.e., $\mathcal{D}_{16}\mathcal{M}(\mathrm{sub.~pix.})$ should be very close to zero.
The reobservation operation \citepalias{Li23} simplifies the map-making procedure by fixing the noise model to be that from the final iteration of the \emph{data} run while keeping the same filtering operation, making it both realization-independent and linear.\footnote{Technically, the reobserved map is also solved with the conjugate gradient method. 
Therefore, the numerical errors in the converged map will have a small fluctuation about the true linear solution. This is found to be negligible for our results.}
This approximation assumes that the template iteration can converge to the true noise model and is independent of the sky signal, which has been tested to be valid at a few percent level for the 40\ghz analysis \citepalias{Li23}. 
We revisit this bias specific to the 90\ghz analysis in Section~\ref{sssec:map-bias}.
The last line in the equation defines the pixel-space transfer matrix $\mathbf{F}_{\mathrm{16},N_F}$ that filters an input map at $N_\mathrm{side}=N_F$ and downgrades it to $N_\mathrm{side}=16$.

\begin{figure}[t!]
    \centering
    \includegraphics[width=\linewidth]{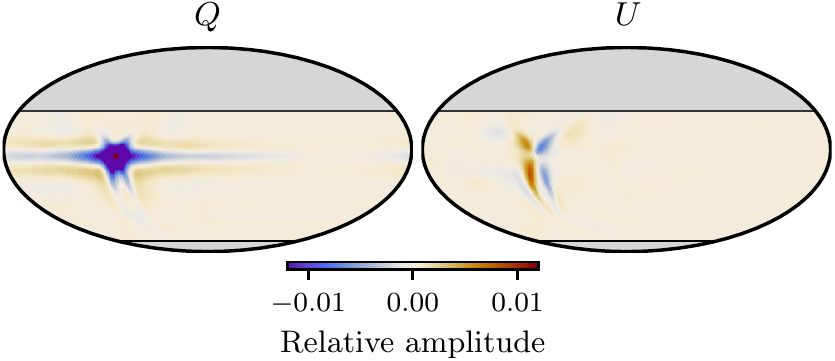}
        \caption{
            Example of the transfer matrix row elements (at native $N_\mathrm{side}=256$ before the downgrade operation) corresponding to the filtering of a single $N_\mathrm{side}=16$ pixel in the Stokes $Q$ map.
            The peak value in the filtered map (among the central red pixels) is $0.94$.
            The ``cross'' shape aligns with the scanning pattern and reflects the finite bandwidth of the filter, while the horizontal stripe arises from the finite timescale over which the filter is subtracted.
            The color scale is highly saturated.
        \label{fig:TF-eg}}
\end{figure}

In principle, the transfer matrix can be analytically constructed by tracking each operation in the pipeline and chaining them in matrix form \citep{BKpure}. 
However, this approach is computationally unfeasible for maximum-likelihood map-making due to the dense noise weighting matrix.
Instead, leveraging the linearity of the reobservation operation, the transfer matrix is constructed by reobserving and downgrading an ensemble of input maps, each with a single pixel set to unity and all others set to zero, and arranging the output maps as column vectors to form a matrix.
This is similar in spirit to other simulation-based approaches \citep[e.g.,][]{BKpure,leung22} but operating in map space.
Figure~\ref{fig:TF-eg} shows an example of the result from an input map with a single Stokes-$Q$ pixel at $N_\mathrm{side}=16$ set to unity: the ``cross'' shape aligns with the scanning pattern and reflects the finite bandwidth of the filter (i.e., the number of harmonic modes in each filter) and the horizontal banding is due to the finite timescale over which the filter is estimated and subtracted (i.e., the number of sweeps of each filter).
Downgrading and concatenating these two maps (Stokes $Q$ and $U$) forms a column in the transfer matrix $\mathbf{F}_\mathrm{16,16}$.

\begin{figure*}[t!]
    \centering
    \includegraphics[width=\linewidth]{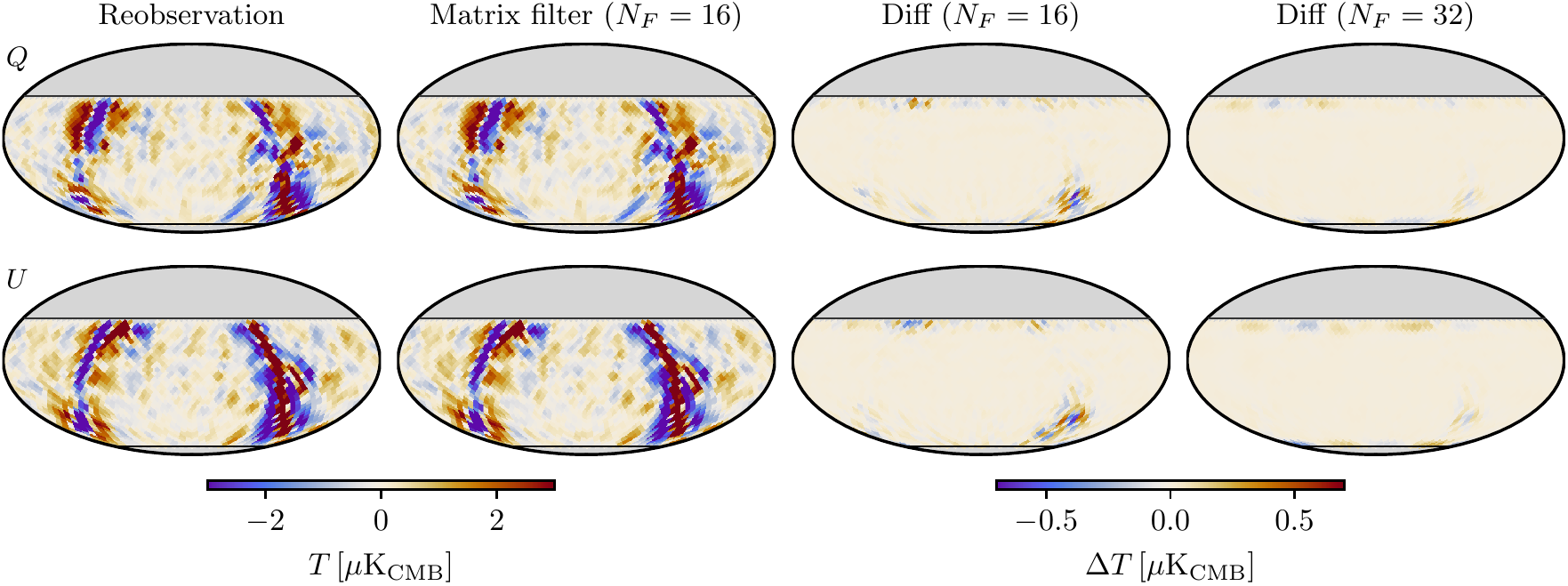}
        \caption{Comparison of reobservation and the transfer matrix approximation. 
        The first column shows the $Q/U$ maps from reobserving the Planck 100\ghz maps and downgrading them to $N_\mathrm{side}=16$. 
        The second and third columns show the same maps but processed with the transfer matrix evaluated with $N_F=16$ ($\mathbf{F}_{16,16}$) and its difference from the reobserved maps (first column). 
        The last column shows the residual maps for a transfer matrix evaluated at $N_F=32$.
        The small discrepancies in both cases present near the pixel scale and localized in regions that are avoided for angular spectrum analysis (Figure~\ref{fig:masks}), and they do not impact the analysis at large angular scales ($\ell \leq 30$).
        \label{fig:reobs-compare}}
\end{figure*}

Figure~\ref{fig:reobs-compare} compares the efficacy of the transfer matrix approximation with the true reobservation for matrices evaluated at $N_F=16$ and $N_F=32$. The residuals in the last two columns show features mainly around the edge of the survey boundary and the Galactic plane. 
The former is due to mode coupling caused by spherical harmonic transformations over a cut sky, and the latter is likely the result of non-zero aliasing from the map-making. 
The omitted sub-pixel signal in Equation~\ref{eq:tfpix-approx-2} could indeed be moved to larger scales or simply shifted outside the original pixel location and show up as residuals. This effect happens mostly around regions in the sky with high signal gradients and can be mitigated with higher $N_F$. 
Nevertheless, this effect, even for $N_F=16$, does not statistically alter the power spectrum estimation at the targeted range $\ell<40$ as shown in the validation in Appendix~\ref{app:xqml} (Figure~\ref{fig:xqml-validation}).

The transfer matrix serves two purposes in this analysis: 1) to forward model the filtering effect, which is useful for processing external template maps and their noise covariance matrices for foreground reduction, and 2) the matrix can be used directly in a quadratic power spectrum estimator to correct for filtering bias in a near-optimal way.
The implementation details of the latter are discussed in Appendix~\ref{app:xqml}.
In practice, in order to apply the matrix to an input map, the map must first be downgraded (with $f_\ell$ from Equation~\ref{eq:fl}) to $N_\mathrm{side}=N_F$ to match the matrix's dimensions. 
This would cause a double counting of the anti-aliasing filter,  as the filter matrix inherently includes $\mathcal{D}_{16}$ (Equation~\ref{eq:tfpix-approx-3}). 
To avoid this, we evaluate the transfer matrix at $N_F=32$, ensuring that the additional $f_\ell$ for $N_\mathrm{side}=32$ has no effect within the target range $\ell\leq32$ due to the filter design (Equation~\ref{eq:fl}). 
Since covariance matrices at $N_\mathrm{side}>16$ resolution for external products are not available, we use $\mathbf{F}_{16,16}$ for noise covariance matrices filtering but apply a window function correction in the harmonic space.

\subsubsection{Isotropic Transfer Functions}
For power spectrum analysis at intermediate angular scales ($\ell>30$), we also model the transfer functions in the traditional way:
\begin{equation}
    \hat{C}_\ell = F_\ell C^\mathrm{in}_\ell,
\end{equation}
where $\hat{C}_\ell$ is the power spectrum of the filtered map computed using \polspice \citep{polspice} and $C^\mathrm{in}_\ell$ is the true power spectrum.
This is a simplification of the treatment presented in \citetalias{Li23} by ignoring mode coupling (i.e., $F_\ell$ is a multiplicative factor for each multipole) in addition to the isotropic filtering approximation. 
This approximation is sufficient in the $\ell>30$ regime, where $F_\ell$ will be exclusively used.
This isotropic transfer function was estimated from the power spectra ratio of the reobserved maps and that from the simulation inputs.
Figure~\ref{fig:filtering_tf} shows $F_\ell$ in green.
Although the transfer function is presented down to $\ell=2$, caution should be used when interpreting the results at $\ell < 30$ as mode coupling is ignored. 
For example, $F_\ell$ is essentially zero for $\ell<5$ due to a combination of mode coupling and numerical accuracy. 
However, as is shown in Section~\ref{sec:tau} and Appendix~\ref{app:xqml}, the information at these scales can still be recovered reliably in the cross-spectrum with the quadratic estimator.

\subsubsection{Maximum-likelihood Mapmaking Bias}\label{sssec:map-bias}
As explained in Section~\ref{sssec:pix-tf}, all transfer functions used for this work assume linearity of the map-maker and are estimated from {\em reobservation} simulations that enforce the map-making weights to be realization-independent.
\citetalias{Li23} showed for the CLASS 40\ghz data that, in the absence of time-domain filtering and for a synchrotron-like signal spectrum the bias in the power-spectrum is $\lesssim 2.5\%$ for $\ell\leq5$. 

For the 90\ghz analysis, we performed an end-to-end simulation to characterize this bias in the presence of filtering.
The simulation takes as input the Planck FFP10 fiducial foreground maps \citep{planck2016-l03} at 100\ghz and Monte Carlo realizations of the $\Lambda$CDM $E$-mode-only simulations \citep{planck2016-l05}.
These sky realizations were projected to the demodulated TOD with realistic 90\ghz noise properties and were processed through the map-making pipeline, including the time-domain filtering and the noise model estimation at each template iteration.
Since we are only interested in the effect on the CMB component, at the last iteration step, after the noise model was re-estimated, the foreground and noise components were offloaded, and only the Monte Carlo CMB component was mapped.
The result of this simulation differs from reobservation in that the effect of realization-dependent weighting (i.e., the non-linearity) is taken into account and a comparison with the reobserved map would reveal this bias.

\begin{figure}[t!]
    \includegraphics[width=\linewidth, ]{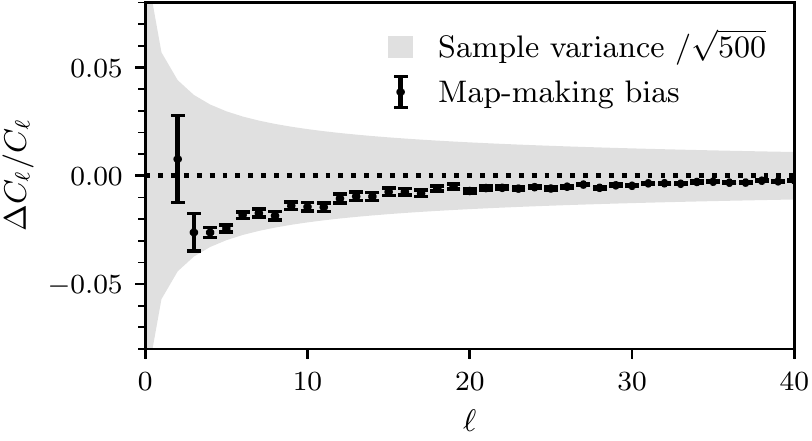}
    \caption{
    Low-$\ell$ bias from the maximum-likelihood map-making noise weighting.
    The error bars are derived from Monte Carlo simulations and represent the uncertainty on the mean.
    The gray shade is the theoretical calculation of the sample variance on the mean, which is larger than the simulation uncertainties due to the canceling effect in the construction of the transfer function. 
    }
    \label{fig:map-bias}
\end{figure}

We used the same low-$\ell$ power spectrum estimation configuration for spectral comparison as the CLASS $\times$ Planck analysis (more details in Section~\ref{ssec:low-ell-spec}). 
Specifically, we used a quadratic cross-spectrum estimator \citep[xQML; adapted from][]{xQML} to cross the simulated maps and the input (unfiltered) CMB realizations over $50\%$ of the sky (\texttt{f050}, Figure~\ref{fig:masks}) with transfer matrix correction (Appendix~\ref{app:xqml}).
The resulting spectra were compared to those from the reobserved maps, both processed through the same pipeline. 
The fractional ratio is shown in Figure~\ref{fig:map-bias}, which eliminates the filtering bias\footnote{While our implementation of the xQML estimator already accounts for filtering bias, this ratio remains effective in reducing common sample variance.} and clearly reveals the bias from noise weighting.
The maximum of this bias is found to be less than $3\%$ at $\ell=3$, consistent with the findings in \citetalias{Li23}. 
However, whereas \citetalias{Li23} analyzed CLASS internal cross spectra, our estimate is from cross-correlation with an unbiased map; i.e., the bias is actually more substantial in the 90\ghz maps, likely due to the smaller frequency-bin sizes chosen for the updated noise model.
The precision of the bias characterization here is significantly improved \citepalias[cf.][]{Li23} due to the TOD-level foreground and noise components removal, which reduces the associated uncertainties.
The shaded regions in Figure~\ref{fig:map-bias} represent the $1\sigma$ uncertainty due to the sample variance of 500 simulations. 
Notably, this uncertainty is much larger than the scatter in the ratio curve because the sample variation present in both the simulated and reobserved maps largely cancels out.
For comparison, the validation plot shown in Figure~\ref{fig:xqml-validation} in Appendix~\ref{app:xqml} is between the reobserved map and the input. 
Therefore, the product of the two (where the uncertainty in the latter is dominated by sample variance as shown in Figure~\ref{fig:xqml-validation}) gives the residual end-to-end bias from the full pipeline after bias correction, including the potential error from the transfer matrix correction through xQML, which assumes linearity of the map-maker.

Given its non-linear nature and the low value estimated for the bias (since the low-$\ell$ spectrum amplitude scales as $\tau^2$, the effect on $\tau$ would be half as large), we do not apply this correction to the CMB power spectra, but the bias in the foreground component and its impact on downstream foreground cleaning is correctly modeled with simulations (see Section~\ref{ssec:simulations}).

\section{Data Validation}\label{sec:validation}
\subsection{Internal Consistency Test}
Potential systematic errors in the data and the validity of the CLASS simulation pipeline were checked through a series of null tests.
The methodology follows the approach outlined in \cite{eimer23}, which we briefly summarize here while highlighting several key differences.

\begin{figure}[t!]
    \centering
    \includegraphics[width=\linewidth]{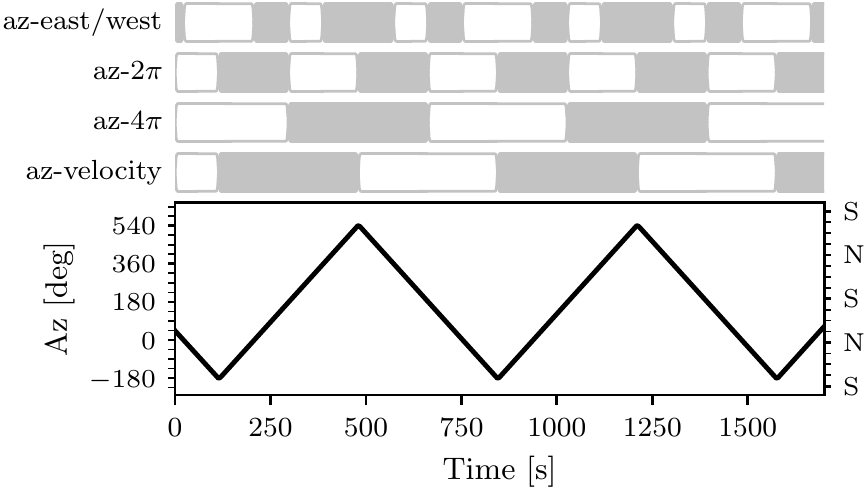}
        \caption{Definition of the scan-related data splits.
        The bottom panel shows the typical $2\,\mathrm{deg\,s^{-1}}$ azimuth scan pattern of the CLASS telescope over two back-and-forth scanning periods (four sweeps). The four az-related splits divide every scan period into two halves and are displayed as boxes with alternating shades. 
        The \texttt{az-east/west} and \texttt{az-velocity} splits are based on the azimuth pointing and velocity directions respectively.
        The \texttt{az-2$\pi$} and \texttt{az-4$\pi$} splits are defined by their azimuth intervals.
        \label{fig:az-splits}}
\end{figure}

\begin{figure}[t!]
    \centering
    \includegraphics[width=\linewidth]{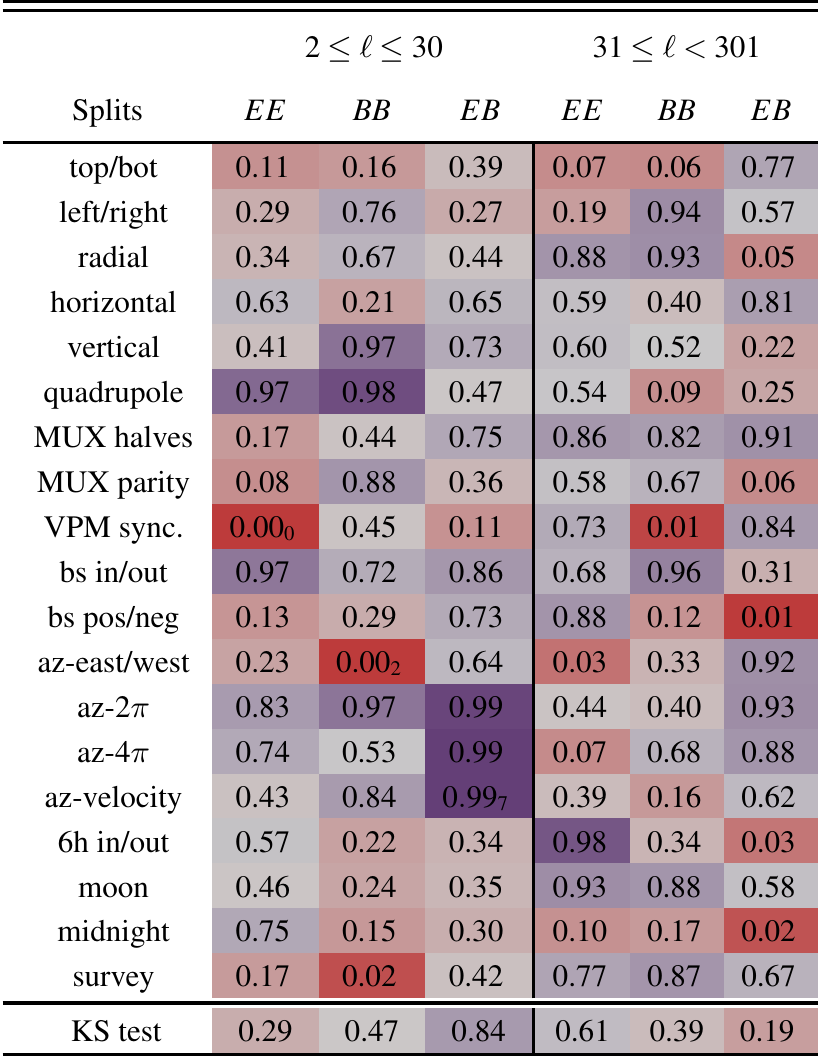}
        \caption{Probability-to-exceed (PTE) values for the low-$\ell$ and mid-$\ell$ null tests.
        The PTE values are tabulated for every null test split and each of the three linear polarization spectra for the two $\ell$-ranges.  
        The cell color indicates the PTE value---red (purple) for PTE values less (greater) than 0.5, meaning that the simulation shows less (more) scatter than the data.
        Color saturation scales with the extremeness of the PTE (deviation from $0.5$).
        A subscript third significant digit is included in cells otherwise rounding to 1 or 0.
        The last row lists the one-tailed p-value from a Kolmogorov Smirnov (KS) test assessing the uniformity of PTE values across splits.
        \label{fig:null-pte}}
\end{figure}

Null maps were generated by differencing the maps created from two splits of the data (null test splits), denoted as A and B, which may be susceptible to different levels of systematic errors. The definitions of these data splits are nearly identical to those detailed in \cite{eimer23}. 
Broadly speaking, 9 splits (the first nine in Figure~\ref{fig:null-pte}) are defined based on detector locations on the focal plane, aiming to test systematics related to optics and geometry, VPM effects, detector crosstalk, readout wiring, atmospheric emission, etc. 
The definitions are the same as those for the 40\ghz data but customized for the multi-module 90\ghz receiver \citep{datta23}. 
The remaining 10 splits are temporal, dividing the data into two halves at different time periods to probe scan-related systematics, telescope pointing error, side-lobe pickup from the Moon, and secular drifts in the instrument over scales ranging from the scanning period ($\sim10$ minutes) to the survey length (years).
Among this set, four splits targeting specifically the scan-correlated effects are illustrated in Figure~\ref{fig:az-splits}. 
The \texttt{az-east/west} checks the consistency of data with completely different ground signals; the \texttt{az-velocity} split probes anything that might be caused by the azimuth servo motor; the \texttt{az-4$\pi$} split (previously denoted \texttt{half sweep} in \citealt{eimer23}) covers the turnaround point with different acceleration directions and is complimented by the new \texttt{az-2$\pi$} split that covers periods preceded or followed by a telescope turnaround.
Due to the limited temporal coverage of the current survey, we did not perform null tests for all instrument configuration changes listed in Figure~\ref{fig:class-overview}. 
Instead, we rely on the ``survey" split to probe potential systematics evolving over time. This split divides the data at 2023-05-23, approximately coinciding with the elevation change and the cage separator installation.

Since filtering operations might be different between two null test splits, the ``null map'' is not necessarily signal-free but may contain residual (differential) filtering artifacts. 
This effect was simulated by reobserving a fiducial sky map from the \sroll 100\ghz map \citep{delouis19} for each split, with the resultant template subtracted for each null map \citep[see Fig. 4 of ][for an example]{eimer23}.
Due to the difference in instrument bandpasses and residual systematics in Planck 100\ghz maps \citep{planck2016-l03,delouis19}, the reobserved maps do not fully capture the filtering effect close to the Galactic plane.
However, this is not a concern for the null test where the cross-spectra were computed with the \texttt{f070} Galactic mask (Figure~\ref{fig:masks}, bottom-left panel).

For each null split, four time-interleaved base maps with evenly assigned spans were made, balancing instrument configuration and environmental conditions.
For each null test, pair-difference maps were created between the two splits for each base map, and cross-spectra between all pairs of base maps were then calculated and averaged over.
This improves upon the previous analysis \citep{eimer23}, which only computed a single cross-spectrum between two base maps. 
The average of all four-way-split cross spectra here is more sensitive to potential systematic issues.
Finally, the statistics of these null cross spectra were compared against a suite of noise-only simulations to assess consistency.
We created 70 simulations from the demodulated TOD level, processed them through the same pipeline, and made null maps for each null test and base split. 
When calculating the null cross-spectra, we mixed and matched the simulation index between pairs of base splits to create an ensemble of 4900 null spectra to compare with the data \citep{eimer23}. 
Spectra for both the data and simulations were calculated with \polspice \citep{polspice} without bias correction.
Mode couplings due to cut sky were accounted for in the covariance matrix computed from the simulation spectra.

To test the internal consistency of the data, we calculated the probability-to-exceed (PTE) values for two angular ranges: The low-$\ell$ set includes single multipoles from $\ell=2$ to $30$, and the mid-$\ell$ set includes $\ell=31 \text{--} 301$ with $\Delta \ell=10$ binning.
Figure~\ref{fig:null-pte} shows the PTE values for all tests. 
The last row of the table records the single-tailed p-value (equivalent to PTE defined above) of the Kolmogorov–Smirnov (KS) test to check if PTE values among splits for each spectrum are consistent with a uniform distribution---all p-values are within the range $0.05-0.95$, indicating good consistency.
The lowest PTE value across the board is found for the low-$\ell$ $EE$ of the \texttt{VPM sync} split, with a value less than $2\times10^{-4}$. 
This split was designed to probe halves of detectors subjected to lower-/higher-than-average VPM synchronous signals \citep{harrington2018thesis,eimer23}. 
This is considered a failure--- even accounting for the look-elsewhere effect \citep{gross10} of examining $6\times19=114$ spectra, the probability of finding such an extreme PTE is less than $0.2\%$.
Upon closer inspection of the null spectra, we found that this is mostly driven by a single multipole at $\ell=8$. 
We conclude that the low PTE is likely caused by inadequate noise modeling under edge conditions with unpaired detectors, which compromises the instrument’s ability to effectively cancel systematics between detector pairs, rather than the VPM emission that this split was originally designed for.
Further discussion of this particular data split is given in Appendix~\ref{app:nulltest}.
We also verify in the subsequent analysis that this single failure does not impact the cross-correlation results.

\subsection{Consistency with Planck}\label{ssec:external-validation}
Given the focus on the low-$\ell$ modes in this work, we do not attempt a cosmological parameter inference from the intermediate angular scales. 
Instead, we used the external Planck data to validate the consistency of the CLASS measurement with $\Lambda$CDM and to obtain an absolute polarization amplitude calibration.

\begin{figure*}[t!]
    \centering
    \includegraphics[width=\linewidth]{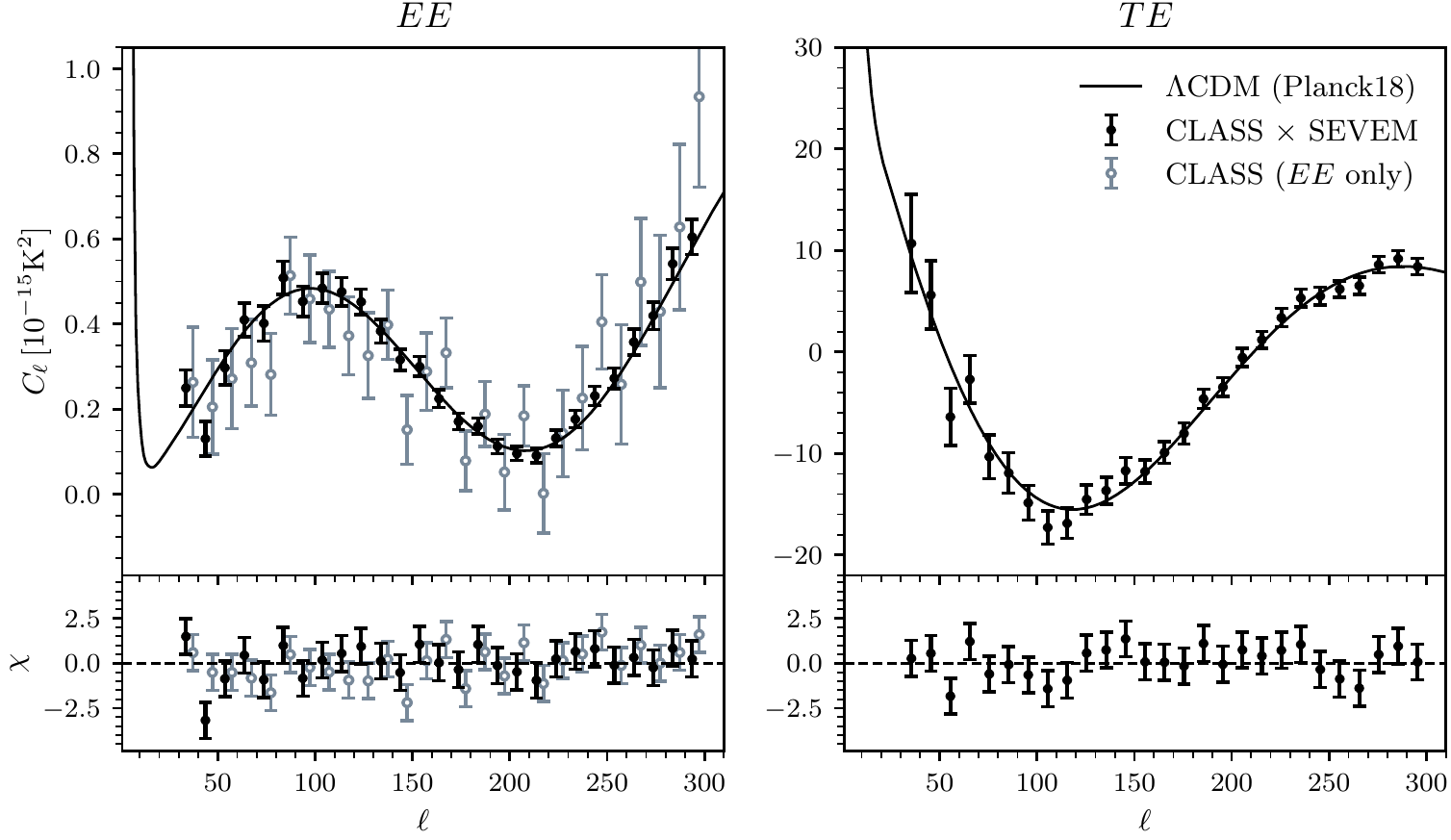}
        \caption{
            $E$-mode power spectra from the CLASS 90\ghz and Planck data.
            The black data points are cross-spectra between CLASS 90\ghz and Planck PR4 \texttt{SEVEM} CMB maps for $EE$ and $TE$ cross-correlations.
            The $TE$ spectrum on the right panel uses the CLASS $E$-mode map and the Planck temperature map.
            The error bars are derived from the diagonal elements of the covariance matrix estimated from 500 simulations including the sample variance.
            The gray open-circle data points in the left panel represent the $EE$ power spectrum from the CLASS 90\ghz data alone, with foreground reduction applied using external data sets. This spectrum is obtained by averaging over six combinations of cross-spectra derived from four time-interleaved map splits.
            The fiducial $\Lambda$CDM spectra based on the Planck best fit are shown as the black curve for comparison.
            The CLASS maps have been adjusted for a $1.053$ absolute calibration factor.
            The bottom panels show the residuals of the calibrated spectra from the Planck best fit.
        \label{fig:spec-high}}
\end{figure*}

To isolate the CMB signal, we used the \texttt{SEVEM} CMB component map from the Planck PR4 \footnote{We used Planck PR4 mainly for its improved systematic error control on large angular scales in polarization.} data release \citep{planck2020-LVII} to cross-correlate with the CLASS 90\ghz map.
The cross-correlation was performed using \polspice with the same mask used for the null test (Figure~\ref{fig:masks}) that has 70\% retained sky fraction after Galactic avoidance.
The estimated power spectra have been corrected for the beam and pixel window functions of both maps and are further corrected for the demodulation low-pass filter for the CLASS map.
Finally, the agreement with the model is checked by the following $\chi^2$ statistic:
\begin{equation}
    \chi^2 = \eta^{-2}(\eta\hat{C}_b-F^{1/2}_b C_b)\mathbf{\Sigma}^{-1}_{bb^\prime}(\eta\hat{C}_{b^\prime}-F^{1/2}_{b^\prime} C_{b^\prime})^T,
\end{equation}
where $\hat{C}_b/C_b$ denote band power from the estimated and fiducial spectra \citep{planck2016-l06} defined between $31\leq \ell<301$ with $\Delta\ell=10$ bins, $\eta$ is the absolute calibration factor,\footnote{The calibration factor correction to the noise covariance matrix is appropriate when the CLASS $E$-mode power spectrum is noise dominated. This procedure has been tested to give unbiased estimation within uncertainties.} $F^{1/2}_b$ is the CLASS mapping transfer function for cross spectra,\footnote{Here we took the square root of the auto-spectra transfer function presented in Figure~\ref{fig:TF-eg}. 
Strictly speaking, this is not necessarily the same as the transfer function for the cross-spectra between CLASS and Planck due to filtering anisotropy and different sky masks over which the spectra are taken. However, we found negligible differences. } and $\mathbf{\Sigma}$ is the band-power covariance matrix.
The estimated band power $\hat{C}_b$ has also been corrected for the beam and pixel window functions.
The $E$-mode transfer function of the Planck map from the PR4 processing \citep{planck2020-LVII} is negligible at the scales of interest.
Here, the covariance matrix was estimated from simulated spectra by pairing 500 CLASS and SEVEM noise simulations with random CMB realizations drawn from the fiducial $\Lambda$CDM model.
The covariance matrix was further corrected by the Hartlap factor \citep{hartlap07} of $1.12$ to account for the finite number of simulations.
A minimization of the $\chi^2$ with respect to $\eta$ gave the best-fit calibration factor $\eta=1.053\pm0.014$. 
The fit is driven by the $EE$ spectrum due to the higher sample variance of the $TE$ spectrum from the Planck temperature map.
The calibration-adjusted spectra have a $\chi^2=53.89$ with respect to the fiducial CMB spectra for $54$ degrees of freedom, or an equivalent PTE value $0.48$.
Figure~\ref{fig:spec-high} shows the adjusted spectra, along with the residual differences expressed in $\chi$.
Of note, the absolute calibration leverages the CMB power spectra at scales $\ell>30$, which is proportional to $A_\mathrm{s}e^{-2\tau}$. 
Therefore, it does not compromise the experiment's ability to measure $\tau$ independently of the calibrator at large angular scales ($\ell<30$), to the extent that $A_\mathrm{s}e^{-2\tau}$ is considered a well-constrained parameter.

As an additional consistency check, we estimated the CMB $E$-mode spectra using CLASS~90\ghz data alone, with foreground reduction applied using external data sets. 
The foreground cleaning procedure follows the approach detailed in Section~\ref{ssec:foreground} for low-resolution maps at $N_\mathrm{side}=16$. 
Since the foreground power dominates over large angular scales, it is a valid approximation to use these results for estimating the CMB at intermediate angular scales.
Two sets of foreground templates were used: the WMAP $K$-band \citep{bennett13} and the Planck 30\ghz map from the PR4 processing \citep{planck2020-LVII} for synchrotron, and two data splits of the Planck 353\ghz map \citep{delouis19} for dust. 
The foreground cleaning coefficients and their uncertainties are summarized in Table~\ref{tab:ilc} denoted as ``CLASS-K'' and ``CLASS-PR4''.
To build a near-optimal spectrum estimation free of noise bias, we used four time-interleaved map splits and computed the cross-spectra between all six non-repeating pairs of maps. 
Foreground cleaning was applied separately for each pair of maps using the predetermined coefficients for the two sets of templates. 
The resulting foreground-cleaned CMB $E$-mode spectrum is shown as gray open circles in the left panel of Figure~\ref{fig:spec-high}. 
Error bars are derived from simulations of the four map splits, with uncertainties from sample variance and foreground cleaning residuals consistently propagated.
After applying the aforementioned calibration factor determined from cross-spectra with Planck, we found $\chi^2 = 25.94$ for 27 degrees of freedom, corresponding to a PTE of 0.52 relative to the Planck best-fit cosmology \citep{planck2016-l06}.

\section{Measurement of Reionization Optical Depth}\label{sec:tau}
Despite efforts to correct the filtering effect, the current signal recovery of the CLASS 90\ghz data is insufficient to carry out an independent measurement of the low-$\ell$ spectrum. 
Instead, we present a cross-correlation analysis with Planck maps to demonstrate the current capability of the CLASS 90\ghz data.
The method follows closely the simulation-based likelihood built for Planck HFI cross spectrum analyses (\texttt{SimAll}; \citealt{planck2014-a10,planck2016-l05,planck2020-LVII}; \citetalias{pagano19}).

\subsection{Simulations and Noise Covariance Matrices}\label{ssec:simulations}
Map simulations are needed for both the foreground cleaning and likelihood modeling in this analysis.
For \sroll maps at 100/143 and 353\ghz, we used their official 500 end-to-end simulations that take a fixed sky realization \citep[CMB and foregrounds;][]{planck2016-l03} as input to the processing pipeline. 
The \sroll release removed the input signal in the end to form realistic noise-plus-systematics simulations.
These signals were added back for the analysis here.
The PR4 30\ghz map also comes with end-to-end simulations in the same style, and we used their realizations $200-699$ to pair with the \sroll ensemble.
For WMAP frequency maps, where only the noise covariance matrices are available, noise realizations were drawn from the noise covariance matrix and added with the Planck fiducial CMB and foreground map.
Here we used the FFP10 fiducial foreground template built for 30\ghz \citep{planck2016-l03} and scaled it by a factor of $2.1$ to roughly account for the band center difference between WMAP $K$-band and Planck 30\ghz. 
Since we care mostly about the scatter among the simulations rather than their mean, the exact scaling factor is not critical.
When these maps were used for cleaning the CLASS data, we used the transfer matrix $\mathbf{F}_{16,32}$ introduced in Section~\ref{sssec:pix-tf} to simulate the filtering effect before pairing them with CLASS maps.

CLASS simulations were generated by taking the FFP10 fiducial sky map at 100\ghz as a sky template and adding time-domain noise realizations for full mapping runs. 
The noise models in map-making were re-estimated for each realization, and five template iterations were carried out as the data processing.
Consequently, any residual systematics related to the non-linearity of the map-making process would be captured in this suite and carried forward through subsequent analyses (see Section~\ref{sssec:map-bias}).

Low-resolution noise covariance matrices are needed for optimal foreground reduction and power spectrum estimation. 
For Planck PR4 30\ghz maps and WMAP maps, we used their official data product; for the \sroll version of the Planck maps, we followed the approach in \citetalias{pagano19} and used the suboptimal choice of noise covariance matrices built for Planck PR3 products \citep[FFP8;][]{planck2014-a14}.
CLASS noise covariance matrix was built from noise-only simulations that ignore the systematics introduced by map non-linearity.
All noise covariance matrices were augmented by a $20\,\mathrm{nK}$ regularization white noise at $N_\mathrm{side}=16$ to ensure numerical stability during matrix inversion. 
As with the maps, when the external noise covariance matrices were used with the CLASS map, they were filtered with the transfer matrix $\mathbf{F}_{16,16}$.

\subsection{Foreground Reduction}\label{ssec:foreground}
In addition to the CLASS 90\ghz, we chose the \sroll processing of the Planck HFI 100/143\ghz maps \citep{delouis19} as the CMB channel ($\mathbf{m}_c$) for cross-correlation.
Both Planck and CLASS maps need to be cleaned of synchrotron and dust polarization emission.
Two data splits (ds) of the \sroll 353\ghz map were used as dust templates ($\mathbf{m}_d$) for CLASS and Planck maps.
Due to the lack of a consistent LFI processing from the \sroll data release \citep{delouis19}, we used the WMAP $K$-band \citep{bennett13} and the Planck PR4 version of the 30\ghz map \citep{planck2020-LVII} as synchrotron templates ($\mathbf{m}_s$) for CLASS and Planck maps respectively.
We did not use the previously released CLASS 40\ghz map (\citetalias{Li23}; \citealt{eimer23}) as the synchrotron tracer mainly because of its lower sensitivity to synchrotron signal compared to lower frequency tracers and the complexities involved in accounting for different transfer functions in CLASS maps. 
A map-level joint analysis of CLASS multi-frequency maps with WMAP/Planck to maximize sensitivity is planned for future work.

\begin{figure}
    \centering
    \includegraphics[width=\linewidth]{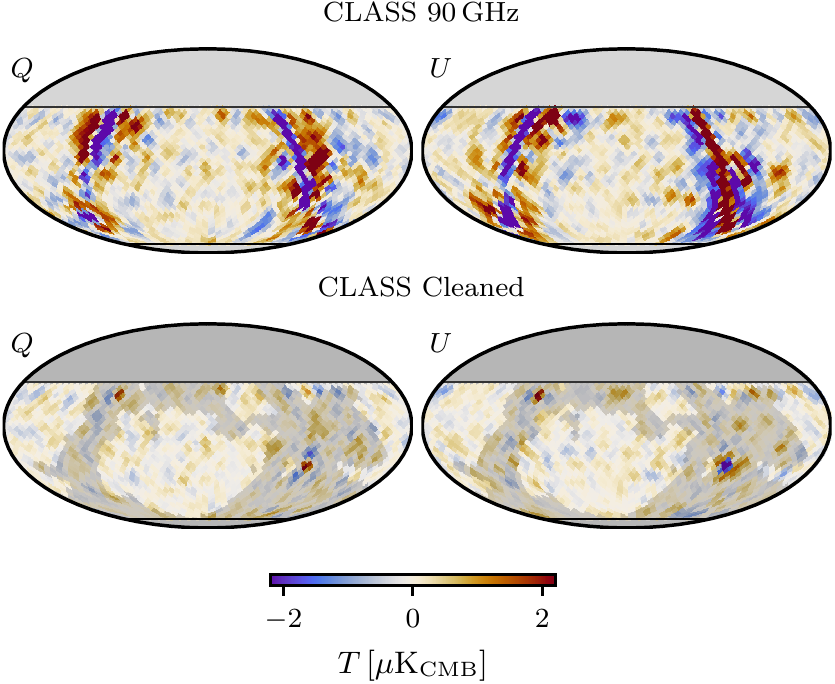}
        \caption{
            Low-resolution CLASS 90\ghz linear polarization maps before (top) and after (bottom) foreground reduction.
            Small residuals appear mostly around the Galactic center and the anticenter (Tau~A) where the data are not used for determining the template coefficients.
            These residuals are not included in power spectrum analysis with even more aggressive masking.
            The baseline \texttt{f050} mask is indicated by the gray shadow in the bottom row.
        \label{fig:map-ilc}}
\end{figure}

The foreground reduced map
\begin{equation}
    \hat{\mathbf{m}}_\mathrm{cmb} = \frac{\mathbf{m}_c -\alpha\mathbf{m}_s -\beta\mathbf{m}_d}{1-\alpha-\beta},
\end{equation}
can be obtained by finding the template coefficients $\alpha$ and $\beta$ through an optimization process that minimizes the following statistics:
\begin{equation}
     \chi^2 = \hat{\mathbf{m}}_\mathrm{cmb}^T \Big(\mathbf{S} + \frac{\mathbf{N}_c +\alpha^2\mathbf{N}_s +\beta^2\mathbf{N}_d}{(1-\alpha-\beta)^2}\Big)^{-1}\hat{\mathbf{m}}_\mathrm{cmb},
\end{equation}
where $\mathbf{S}$ is the fiducial signal covariance matrix that we calculate based on the Planck best-fit parameters \citep{planck2016-l06}; $\mathbf{N}_c$, $\mathbf{N}_s$, and $\mathbf{N}_d$ are noise covariance matrices of maps $\mathbf{m}_c$, $\mathbf{m}_s$, and $\mathbf{m}_d$ respectively.
For the Planck maps at 100 and 143\ghz, the processing is identical to that in \citetalias{pagano19}, where the full resolution maps were downgraded to $N_\mathrm{side}=16$ with $\mathcal{D}_{16}$ and paired with low-resolution noise covariance matrices.
For CLASS foreground cleaning, the template maps (WMAP $K$-band and \sroll 353\ghz \texttt{ds1}) were first downgraded to $N_\mathrm{side}=32$ and then transformed by $\mathbf{F}_{16,32}$ to approximate the reobserved map at $N_\mathrm{side}=16$.
The template coefficients were fit over the sky region defined by \texttt{f070} (Figure~\ref{fig:masks}).

Similar analyses were performed for simulated maps to determine the uncertainties in the template coefficients ($\sigma_\alpha,\sigma_\beta$).
The foreground-reduced simulations were further subtracted by the FFP\,10 fiducial CMB realization (the fiducial CMB realization was filtered for the set used for CLASS map cleaning) to create a realistic simulation suite that encompasses noise and systematics in each frequency map as well as foreground residuals.

The foreground cleaning coefficients for this analysis and several sensibility checks are summarized in Table~\ref{tab:ilc}.
The ``S2'' set aims to reproduce the analysis presented in \citetalias{pagano19} where the \sroll 100/143\ghz maps used WMAP $K$/$Ka$-band as synchrotron templates.
Both maps used the \sroll 353\ghz map for dust removal.
All except for the synchrotron coefficients for 143\ghz are consistent with previous results (\citetalias{pagano19}; \citealt{deBelsunce21}). 
Small differences in the analysis choice (e.g., the masks were defined in celestial coordinates, which would cause differences for boundary pixels at $N_\mathrm{side}=16$) may be driving this discrepancy.
Inspecting the foreground-reduced map, we noticed little difference. 
This is also consistent with the observation that synchrotron cleaning is not critical for Planck 143\ghz maps \citep{planck2016-l05}.
Consistent results were also found when reproducing configurations used in \cite{planck2016-l05} and \cite{deBelsunce21}, reinforcing confidence in our implementation.
The ensuing entry ``S2-PR4'' in Table~\ref{tab:ilc} is the configuration we use to clean both the 100 and 143\ghz maps for the CLASS $\times$ Planck analysis.
This is similar to the S2 set but uses the PR4 30\ghz map as the synchrotron template. 
The dust coefficients are very stable, while the synchrotron coefficients at 100\ghz increase due to the higher band center of the PR4 30\ghz map compared to WMAP $K$-band. 
The synchrotron coefficients at 143\ghz are consistent with those reported in \citep{planck2016-l05,deBelsunce21}, although these studies used 30\ghz maps from the PR3 release.
The foreground-reduced Planck 100 and 143\ghz maps were coadded to cross-correlate with CLASS.

To check the efficacy of performing foreground cleaning on filtered maps, we performed the test denoted as ``S2-Filter'' where the reobserved \sroll 100\ghz map was cleaned with matrix ($\mathbf{F}_{16,32}$) filtered templates from WMAP $K$-band and \sroll 353\ghz. 
It should be noted that filtering operations remove large-scale modes from the map; therefore, the template coefficients are expected to be determined from different angular scales and are not necessarily the same as the unfiltered results. 
The coefficients in Table~\ref{tab:ilc} show a substantial reduction in synchrotron coefficients, albeit with increased uncertainties. 
In contrast, the dust coefficients increased, indicating that more synchrotron signals are preferentially filtered out, consistent with the steeper (redder) angular power spectrum of synchrotron compared to dust \citep{planck2016-l04}.
At the map level, however, the foreground-reduced map from the filtered data is almost identical to the filtered foreground-reduced map (from the S2 configuration) further confirming the validity of the filtered foreground cleaning approach.

Finally, we proceed with the data combination for CLASS data.
The ``CLASS-K'' set is the same as the S2-Filter but swaps out the filtered Planck 100\ghz map with the CLASS 90\ghz map. 
The shifts in coefficients reflect the lower band center of the CLASS 90\ghz array \citep[92\ghz,][]{dahal22}.
The ``CLASS-PR4'' set is similar to CLASS-K but uses the Planck 30\ghz map from PR4 for synchrotron cleaning; the purpose of this combination is to obtain a foreground-cleaned map uncorrelated with CLASS-K noise, intended for the high-$\ell$ internal cross-spectrum analysis in Section~\ref{ssec:external-validation}, but it was not used for the baseline low-$\ell$ likelihood.

The CLASS 90\ghz low-resolution polarization maps at $N_\mathrm{side}=16$ are shown in Figure~\ref{fig:map-ilc} before and after foreground cleaning.  
Although the foreground cleaning coefficients are determined using the conservative \texttt{f070} Galactic avoidance mask, effective foreground reduction is observed even near the Galactic plane. 
This demonstrates that the CLASS measurement of synchrotron and dust at 90\ghz is in good agreement with external foreground tracers.

\begin{deluxetable}{c|cc|cc}
    \tablewidth{\textwidth} 
   \setlength{\arrayrulewidth}{.1em}
    \tablecaption{\label{tab:ilc}
    Template coefficients from the foreground reduction analysis.
    ``S2'' denotes the reproduction of the results from \citetalias{pagano19}, using the exact same data combination (WMAP $K/Ka$ and \sroll 353\ghz);
    ``S2-PR4'' is the same as S2 but uses the PR4 30\ghz map as the synchrotron template for both 100 and 143\ghz maps.
    ``S2-Filter'' uses the S2 configuration but for filtered maps;
    ``CLASS-K'' and ``CLASS-PR4'' are CLASS 90\ghz maps cleaned with the S2 and S2-PR4 configurations, respectively.
    Entries in bold font are configurations used for the baseline low-$\ell$ analysis.
    All numbers correspond to maps in thermodynamic units.
    }
    \tablehead{
        \colhead{} & \colhead{$\alpha\times10^2$} & \colhead{$\beta\times10^2$} & \colhead{$\alpha\times10^2$} & \colhead{$\beta\times10^2$}\\
        \cline{2-5}
        \colhead{{Name}} & \multicolumn{2}{c}{Planck 100\ghz} & \multicolumn{2}{c}{Planck 143\ghz}    
    }
    \startdata
    S2 &  $1.14\pm0.07$ & $1.83\pm0.02$ & $2.58\pm0.21$ & $3.91\pm0.01$ \\
    \textbf{S2-PR4} & $2.26\pm0.14$ & $1.83\pm0.02$ & $1.37\pm0.13$ & $3.92\pm0.01$ \\
    S2-Filter & $0.59\pm0.18$ & $1.96\pm0.02$ &  &  \\
    \colhead{} & \multicolumn{2}{c}{CLASS 90\ghz} & \multicolumn{2}{c}{ } \\
    \textbf{CLASS-K} & $1.24\pm0.08$ & $1.39\pm0.01$ &  & \\
    CLASS-PR4 & $3.43\pm0.17$ & $1.39\pm0.01$ &  & \\
\enddata  
\end{deluxetable}

\subsection{Low-\texorpdfstring{$\ell$}{l} CMB Spectrum}\label{ssec:low-ell-spec}
The covariance matrix of the foreground-reduced maps can be modeled as:
\begin{equation}
    \mathbf{C} = \mathbf{S} + \frac{\mathbf{N}_c +\alpha^2\mathbf{N}_s +\beta^2\mathbf{N}_d +\sigma_\alpha^2 \mathbf{m}^T_s\mathbf{m}_s+\sigma_\beta^2 \mathbf{m}^T_d\mathbf{m}_d}{(1-\alpha-\beta)^2}. \label{eq:fg-red-N}
\end{equation}
To simplify the calculation, we performed a coaddition of the foreground-reduced 100\ghz and 143\ghz maps, with the weights determined from the diagonal component of Equation~\ref{eq:fg-red-N} ignoring the covariance due to common templates.
This coadded map was then cross-correlated with the (foreground-reduced) CLASS 90\ghz map to obtain our final spectrum estimation.

\begin{figure}
    \centering
    \includegraphics[width=\linewidth]{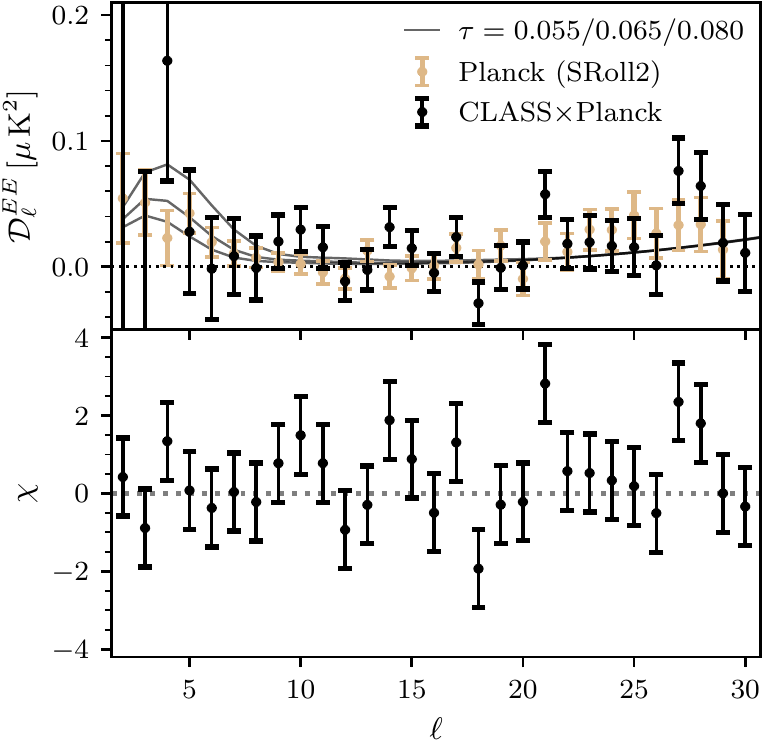}
        \caption{
            Transfer matrix-corrected xQML $EE$ cross spectra between CLASS 90\ghz and the Planck 100/143\ghz coaddition map. 
            The spectra are taken with a baseline \texttt{f050} Galactic mask.
            The Planck $100 \times 143$\ghz result from \citetalias{pagano19} is shown as brown data points for comparison.
            The error bars are derived from a simulation set that includes instrument noise, foreground cleaning residuals, and CMB signals, thereby incorporating sample variance.
            Three theory curves are shown for visual aid for $\tau=0.055/0.065/0.080$ (from bottom to top), roughly corresponding to models preferred by Planck HFI/CLASS (\citetalias{pagano19}; \citealt{planck2020-LVII,tristram24}), Planck LFI/WMAP \citep{lattanzi17,planck2020-LVII,natale20,paradiso23}, and high-$\ell$+CMB lensing \citep{planck2016-l06,giare24}.
            The bottom panel shows the residuals from the $\tau=0.055$ model in $\chi$ values.
        \label{fig:spec-xqml}}
\end{figure}

Quadratic maximum-likelihood estimators \citep[QML;][]{tegmark97,tegmark01,efstathiou06} are optimal for power spectrum estimation \citep{tegmark97,Gerbino20}, particularly on cut-sky and at low-$\ell$.
The idea can be extended to cross-spectra estimation to avoid the need for accurate noise bias subtraction.
Based on a recent implementation by \cite{xQML}, xQML, we further customized the code to allow for the correction of a transfer matrix for either or both maps.\footnote{\url{https://github.com/class-telescope/xQML}}
The matrices listed in Equation~\ref{eq:fg-red-N} and the transfer matrix $\mathbf{F}_{16,16}$ were used to build the xQML estimator.
The details of the algorithm and its validation are presented in Appendix~\ref{app:xqml}.

Figure~\ref{fig:spec-xqml} shows the transfer matrix-corrected xQML cross spectra ($\mathcal{D}_\ell\equiv\ell(\ell+1)C_\ell/2\pi$) between the CLASS 90\ghz map and the foreground-reduced Planck 100/143\ghz coaddition, calculated over the \texttt{f050} mask.
The uncertainty estimates are derived from simulations that include signal realizations at fixed $\tau=0.054$ and foreground-reduced simulations set (with the fiducial CMB realization subtracted); i.e., the error bars include contributions from sample variance. 

\begin{figure}
    \centering
    \includegraphics[width=\linewidth]{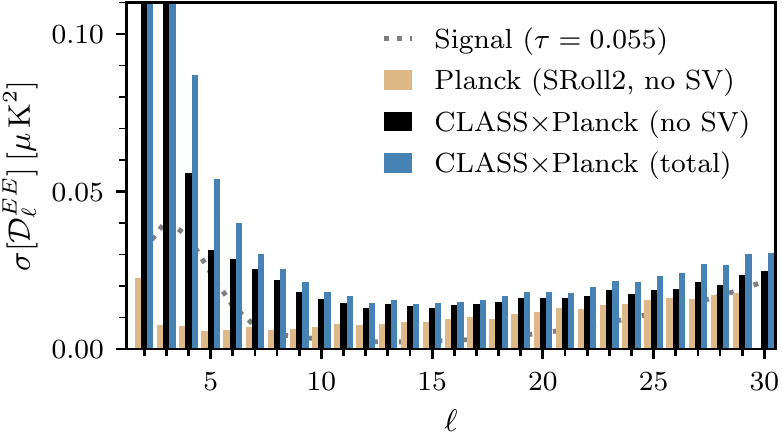}
        \caption{
            Transfer matrix-corrected xQML $EE$ cross spectra error budget. 
            The brown bars show the result from 100$\times$143\ghz analysis \citepalias{pagano19}.
            The black and blue bars show the result for the CLASS $\times$ Planck, without (black) and with (blue) the inclusion of sample variance (SV) for $\tau=0.055$ and the \texttt{f050} mask.
        \label{fig:spec-error}}
\end{figure}

\begin{figure}
    \centering
    \includegraphics[width=\linewidth]{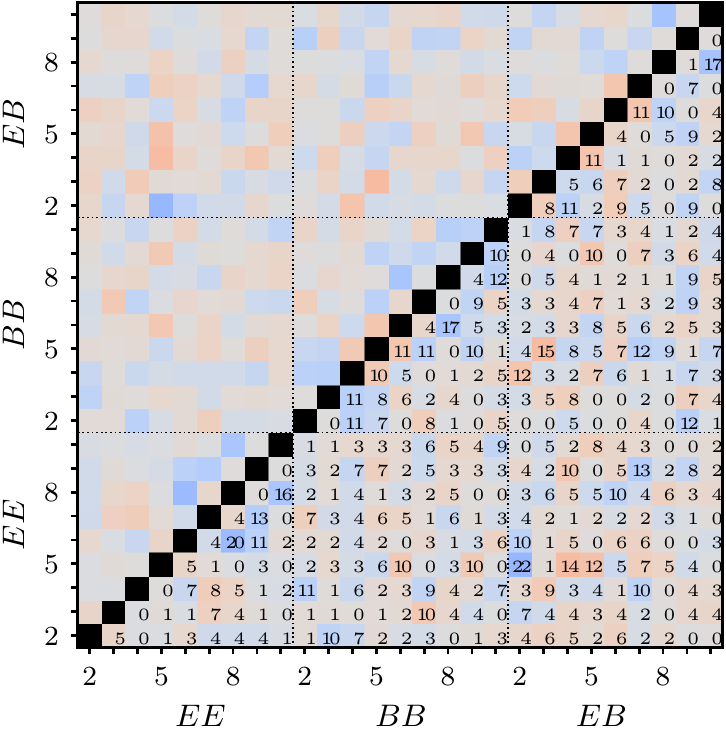}
        \caption{Transfer matrix-corrected xQML spectra correlation matrix ($2\leq\ell\leq10$) for CLASS $\times$ Planck over the \texttt{f050} mask. 
        The matrix is built from 500 simulations combining the foreground-cleaned simulation suite with signals drawn from the fiducial $\Lambda$CDM model.
        The correlation amplitudes are expressed as percentages and labeled in the lower triangle of the matrix.
        Red and blue colors indicate positive and negative correlations, respectively. 
        The diagonal elements are colored black.
        The correlation coefficient amplitudes peak around 20\% on the second off-diagonals, primarily due to the inherent mode coupling from the xQML estimator \citep{xQML}.
        \label{fig:spec-cov}}
\end{figure}

Figure~\ref{fig:spec-error} shows the per-multipole uncertainty from the noise (i.e., not including sample variance, black) and the total error budget (blue, including the sample variance from the fiducial model).
Figure~\ref{fig:spec-cov} shows the correlation matrix of the estimate spectra at $2\leq\ell\leq10$ among all three spectra: $EE$, $BB$, and $EB$.
Given the ensemble size of 500, the off-diagonal elements are expected to have statistical fluctuations at $5\%$ level \citep{lueker2010}. 
The dominant correlations appear on the second diagonal, consistent with the performance of xQML over the cut sky \citep{xQML}. Other significant correlation coefficients are associated with the data and depend on data combinations.
The overall smallness of the off-diagonal correlations suggests that the transfer matrix treatment has largely corrected for mode coupling and polarization leakage due to filtering and the cut sky.

\begin{figure}[t!]
    \centering
    \includegraphics[width=\linewidth]{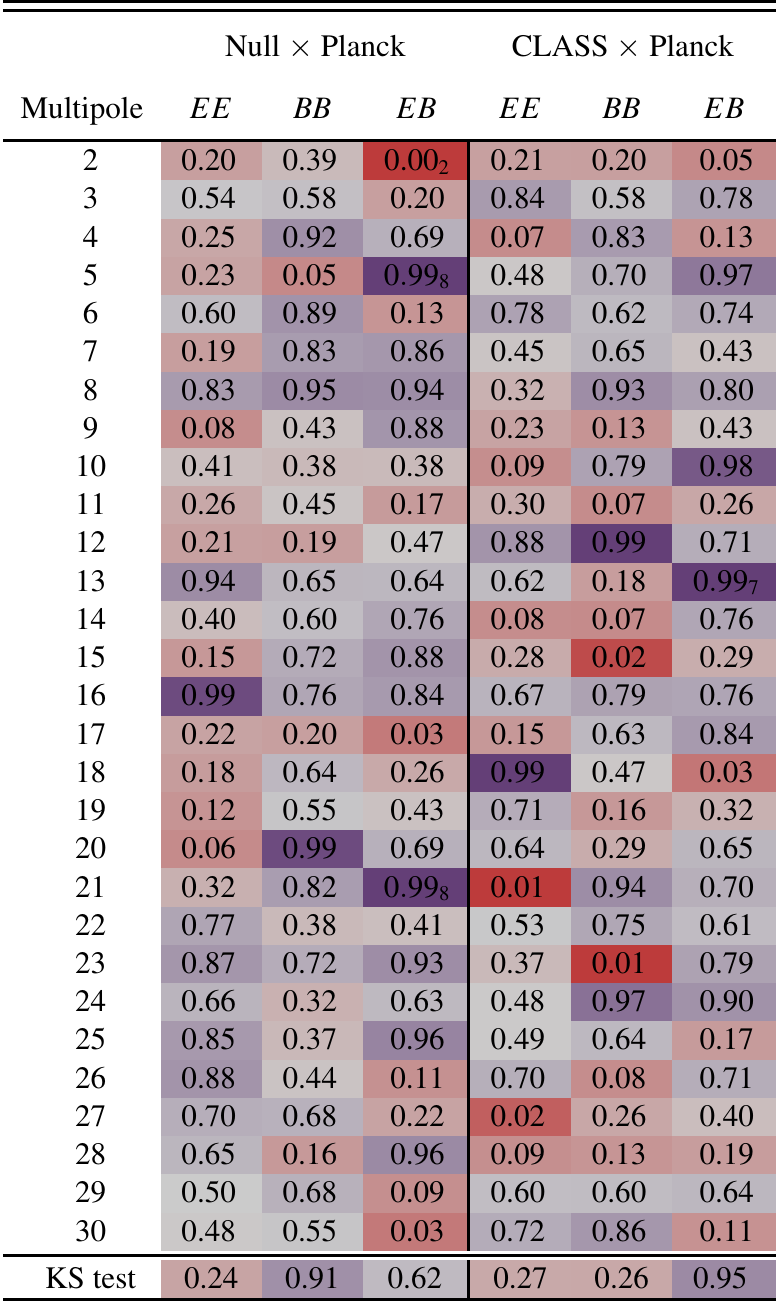}
        \caption{
        PTE values for the Null $\times$ Planck and CLASS $\times$ Planck cross spectra for each multipole $\ell$ and every polarization spectrum. 
        In the former case, the null spectra are compared to the null model,  and in the latter case, the comparison is made against the best-fit model with $\tau=0.053$ (Section~\ref{ssec:low-ell-spec}). Given the limited number of simulations, PTE values are calculated from the cumulative distribution of the simulations using kernel density estimation. The last row lists the one-tailed p-value from a KS test assessing the uniformity of PTE values across multipoles. The color scheme follows that used in Figure~\ref{fig:null-pte}.
        \label{fig:tau-pte}}
\end{figure}

As an additional check, we tested the pipeline for a CLASS null map by differencing two map splits and cross-correlating with the foreground-reduced Planck coaddition map.
The cross-spectra, denoted as ``Null $\times$ Planck'', were compared to an ensemble of simulations.
The simulations for the Planck map contained CMB realizations ($\tau=0.054$) and foreground cleaning residuals as above, but only instrumental noise was included for the CLASS null map.
The per-multipole PTE values are tabulated in Figure~\ref{fig:tau-pte} (first three columns).
The most extreme values are found to be $0.002/0.998$ in the $EB$ spectrum, approaching the precision limited by the number of simulations. 
Considering the total number of tests conducted, these results indicate good overall consistency.

\subsection{Simulation-based \texorpdfstring{$\tau$}{Tau} Inference}\label{ssec:tau-likelihood}
Building on the intuition that the xQML spectra have negligible mode coupling, the likelihood can be written separately \citep{planck2014-a10} for each multipole as:

\begin{align}
    \ln\mathcal{L}(\mathrm{data}|\tau, A_\mathrm{s}) &=  \sum_{\ell=2}^{30} \ln\mathcal{L}(\hat{C}^{EE}_\ell|\tau, A_\mathrm{s}).
\end{align}
where the per-$\ell$ likelihood $\mathcal{L}(\hat{C}^{EE}_\ell|\tau, A_\mathrm{s})$ can be empirically estimated from simulations.
Following largely the steps outlined in \citetalias{pagano19}, 
\begin{enumerate}
    \item 600 pairs of parameters $(\tau, A_\mathrm{s})$ were generated, with $\tau$ drawn uniformly within the  $0$ -- $0.3$ range\footnote{The upper limit is an over conservative choice, mainly to accommodate the forecast simulations in Section~\ref{sec:forecast}.} and $A_\mathrm{s}$ chosen such that $A_\mathrm{s}e^{-2\tau}=1.884\times10^{-9}$. 
    The other cosmological parameters were fixed to the Planck best-fit values \citep{planck2016-l06}.
    \item For each set of parameters, the theoretical power spectra were computed using \texttt{CAMB} \citep{camb}, and 500 CMB realizations were created using the \texttt{Healpix} routine \texttt{synfast}. 
    \item The 500 signal simulations were paired with the foreground-reduced noise simulation suite. For the pairing with CLASS data, the simulated CMB maps were matrix filtered with $\mathbf{F}_{16,32}$.
    \item The xQML spectra were estimated for all 500 simulations, and a kernel density estimation (KDE) was performed for each multipole of the $EE$ spectrum. 
    The KDE amplitude evaluated at $\hat{C}_\ell$ (the data estimate) was used as the likelihood value.
    \item The likelihood values of the 600 pairs of parameters were interpolated as a function of $\tau$ using Gaussian process regression \citep{gpy}.
    \item The final posterior distribution of $A_\mathrm{s}-\tau$ was obtained by Monte Carlo sampling the interpolated likelihood using \texttt{emcee} \citep{emcee}, assuming a positive prior on $\tau$ and a Gaussian prior on $10^9A_\mathrm{s}e^{-2\tau}$ based on the Planck best fit \citep{planck2016-l06}.
\end{enumerate}

The entire pipeline from the foreground reduction through parameter estimation has been checked for the unfiltered S2 configuration, and we found less than $0.2\sigma$ deviation from the \citetalias{pagano19} results under various choices of masking and multipole omission.

\begin{figure}
    \centering
    \includegraphics[width=\linewidth]{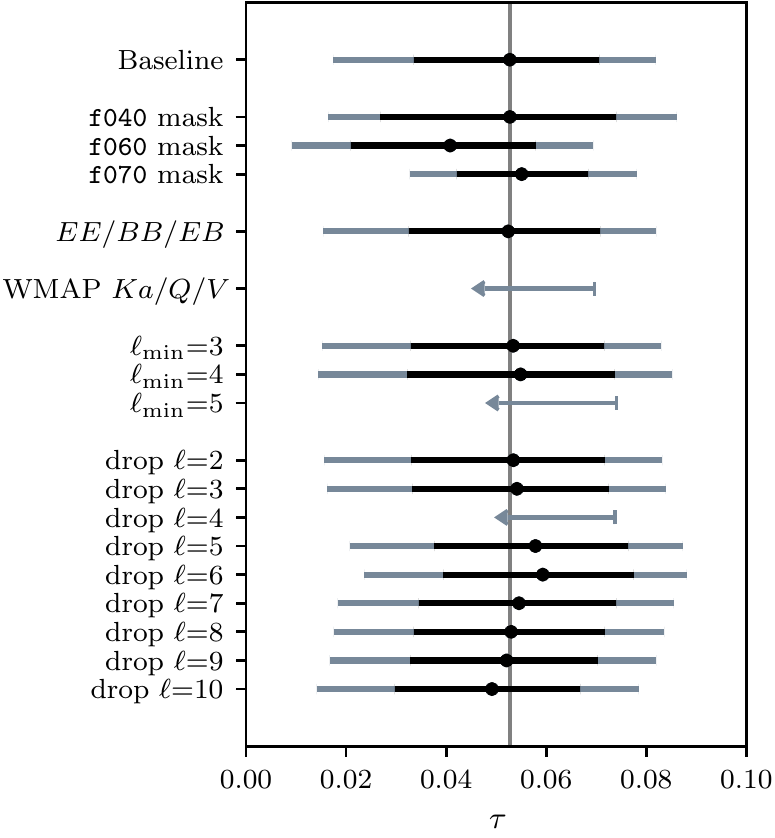}
        \caption{
            Marginalized posterior distribution of $\tau$ from CLASS $\times$ Planck spectra over \texttt{f050} mask starting at $\ell_\mathrm{min}=2$ (baseline) and results from alternative analysis choices including masking, polarization spectra used for likelihood, data set (using WMAP instead of Planck), and multipole selections.
            The black and gray error bars indicate the $68\%$ and $95\%$ confidence intervals respectively.
            The best-fit value for the baseline configuration is also marked by the vertical line.
            In cases where the ``$2\sigma$'' interval (with $1\sigma$ defined by half the $16$ -- $84$ percentile range) extends below $\tau=0$, only the $95\%$ upper limits are shown.
        \label{fig:tau}}
\end{figure}

For our baseline configuration (\texttt{f050} mask and $\ell_\mathrm{min}=2$), the best-fit value (maximum a posteriori estimation) is:
\begin{equation}
    \tau = 0.053 ^{+0.018}_{-0.019}\quad \text{(68\%, CLASS $\times$ Planck)},
\end{equation}
in good agreement with Planck-only results (\citetalias{pagano19}; \citealt{tristram24}).
For the best-fit cosmology, we computed the PTE values for each multipole and spectrum by comparing the data estimate to the distribution of the 500 signal-plus-noise simulations, and the results are listed in Figure~\ref{fig:tau-pte} (last three columns).
Overall, the PTE values are consistent with uniform distributions between $0$ and $1$ across multipoles, as confirmed by the one-tailed KS test for $EE$/$BB$ and $EB$ spectra, suggesting that the model is a good fit to the data.
Since there is no reionization signal present in the $EB$ and $BB$ spectra (assuming no primordial tensor mode) and the CMB lensing signal is negligible, the check on the $B$-mode spectra serves as an additional null test for the analysis pipeline.
Building the covariance matrix for the $EE$ power spectrum over $2\leq\ell\leq30$ from 500 simulations at the best-fit cosmology with appropriate bias correction \citep{hartlap07} yielded $\chi^2=32.7$ for $28$ degrees of freedom, corresponding to a PTE of $0.22$. 
Repeating the likelihood analysis but replacing the CLASS data with simulations fixing $\tau=0$, we found that the best-fit result from CLASS rejects the null hypothesis of no reionization at $99.4\%$ significance.

In addition to the baseline result, we explored alternative analysis choices, including variations in masking, polarization spectra used for likelihood, external data sets, and multipole ranges. 
These $\tau$ constraints are listed in Figure~\ref{fig:tau} alongside the baseline result.
Analysis masks ranging from \texttt{f040} to \texttt{f070} were tested to evaluate the efficacy of the foreground cleaning performed with the \texttt{f070} mask. 
All results are consistent with the baseline within $0.6\sigma$, with the tightest constraint from the \texttt{f070} mask due to the larger sky fraction:
\begin{equation}
    \tau = 0.055 \pm0.013\quad \text{(68\%, CLASS $\times$ Planck, \texttt{f070})}.
\end{equation}
However, we found a notably larger value $\tau = 0.078\pm0.018$ for \texttt{f070} when including all three polarization spectra in the likelihood.
Upon close inspection, this was traced to a $-4.4\sigma$ excursion in the $BB$ spectrum at $\ell=12$.
As is explained below, this is the case where the data favors a larger $\tau$ value to increase the sample variance to better fit the data. This behavior was not seen for other masks (the joint $EE/BB$ and $EB$ likelihood result for \texttt{f050} is also shown in Figure~\ref{fig:tau}), suggesting potential systematics in the extended sky region.
Therefore, we adopt the \texttt{f050} mask and the $EE$-only likelihood as the baseline, following \citetalias{pagano19}.

As an alternative to the Planck HFI maps, we tested WMAP maps as CMB channels for cross-correlation. We cleaned WMAP $Ka/Q$ and $V$ band maps \citep{bennett13} using WMAP $K$ band and \sroll 353\ghz as synchrotron and dust templates, respectively. 
The foreground coefficients are in excellent agreement with previous results \citep{lattanzi17}. The three foreground-reduced WMAP maps were coadded and cross-correlated with the CLASS-PR4 configuration. This setup differs from the baseline result where the CLASS map was cleaned of synchrotron using WMAP $K$-band, but it is more optimal as the WMAP $Ka/Q/V$ channels have lower band centers than CLASS 90\ghz. 
The measurement obtained from this combination over \texttt{f050} yields an upper limit:
\begin{equation}
    \tau < 0.070\quad \text{(95\%, CLASS $\times$ WMAP)}.
\end{equation}

\begin{figure}
    \centering
    \includegraphics[width=\linewidth]{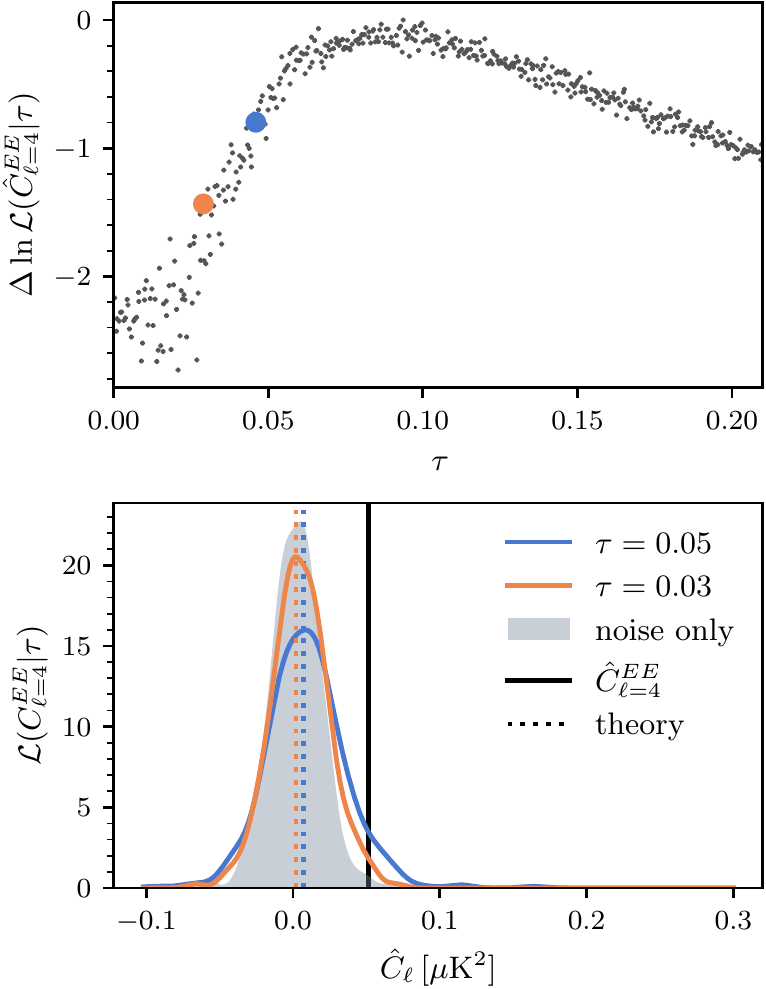}
        \caption{
            \textit{Top}: $\tau$ likelihood for a single multipole, $\ell=4$, of the $EE$ power spectrum evaluated at the measurement $\hat{C}_{\ell=4}^{EE}$. 
            Gray dots are the likelihood values computed for 600 $(\tau, A_\mathrm{s})$ parameter pairs. Blue and orange dots highlight two examples with $\tau=0.05$ and $\tau=0.03$, respectively.
            \textit{Bottom}: The likelihood function is shown for three scenarios: $\tau=0.05$ (blue), $\tau=0.03$ (orange), and a noise-only case (gray). 
            The theory prediction of $C_{\ell}$ in each case is shown as a dotted line with matching colors.
            The vertical black line marks the measured value $\hat{C}_{\ell=4}^{EE}$, where its intersection with each likelihood curve corresponds to the likelihood values depicted in the top panel.
        \label{fig:tau-likelihood}}
\end{figure}

We further checked the robustness of the baseline result by excluding multipoles below $\ell_\mathrm{min}$ or dropping a single multipole at a time.
As is shown in Figure~\ref{fig:tau}, the $\ell=4$ mode plays a pivotal role: omitting this multipole reduces the detection significance below the threshold.
To understand the source of its constraining power, we show in Figure~\ref{fig:tau-likelihood} the simulation-based likelihood for $C^{EE}_{\ell=4}$.  
Two choices of $\tau$ values, $\tau=0.05$ and $\tau=0.03$, are highlighted and their likelihood functions are shown in the bottom panel.
The measurement from data at $\ell=4$ is positive and favors a $\tau$ value significantly higher than $0.05$.
However, the increased likelihood when $\tau$ is raised from $0.03$ to $0.05$ is not solely attributed to closer data--model agreement.
The shift in the theoretical $C_{\ell}$ value (indicated by the vertical dotted line) between these two cases is insufficient to fully explain the observed increase in the likelihood.
Equally important is the increase in the sample variance component in the total variance. 
This effect is reflected by the broader $\tau=0.05$ likelihood function, which raises the likelihood values for data points lying far from the distribution’s centroid. 
Formally, the variance of the cross-spectrum can be expressed as \citep[][Equation 30]{grain12}
\begin{align}\label{eq:variance}
    \mathrm{var} (C_\ell) &= \frac{1}{(2\ell+1)f_\mathrm{sky}}\Big[C^2_\ell + \left(C_\ell+N^\mathrm{C}_\ell/F_\ell\right)\left(C_\ell+N^\mathrm{P}_\ell\right)\Big],
\end{align}
where $f_\mathrm{sky}$ is the sky fraction, $C_\ell$ is the signal spectrum, $N^\mathrm{C}_\ell$ and $N^\mathrm{P}_\ell$ are the noise spectra of CLASS and Planck, and $F_\ell$ is the transfer function (beam window functions are not important here).
Although the fiducial signal is subdominant compared to the noise, it can still appreciably contribute to the total variance through the cross terms.

In summary, the measurement from CLASS $\times$ Planck is fully compatible with the $\Lambda$CDM model plus noise (including instrumental noise, foreground residuals, and modeled systematic uncertainties). However, the constraint is limited by the low-$\ell$ sensitivity of the CLASS map, with the primary contribution coming from the $EE$ spectrum at $\ell$ = 4 through sample variance.

\subsection{Forecast}\label{sec:forecast}
The ultimate goal of the CLASS project is to provide an independent measurement of the reionization optical depth at a precision ($\sigma_\tau$) limited only by the cosmic variance \citep{watts18}.
In this section, we forecast the path to achieve this goal given our understanding of the analysis challenges.

We project two types of improvement to the current data and analysis: 1) relaxing the time-domain filtering, and 2) reducing the noise level. 
For the first aspect, suppose the filtering can be improved by a factor $\lambda$, thereby modifying the pixel-space transfer matrix $\mathbf{F}$ as:
\begin{equation}
    \mathbf{F} \rightarrow \lambda + (1-\lambda) \mathbf{F}.
\end{equation}
Since the retained power at the lowest $\ell$ in the current analysis is essentially zero, $\lambda^2$ is approximately the improved transfer function at $\ell=2$.
The projected map depth improvement is achieved by scaling the noise-only simulations and noise covariance matrices by constant factors. 
This implies an optimistic assumption that the low-$\ell$ red noise scales well with the map depth and does not inflate as the time-domain filtering is relaxed (before applying the transfer function correction). 
These assumptions might not hold for components such as linearly-polarized atmospheric emission \citep{Li23cloud,coerver24}.

\begin{figure}
    \centering
    \includegraphics[width=\linewidth]{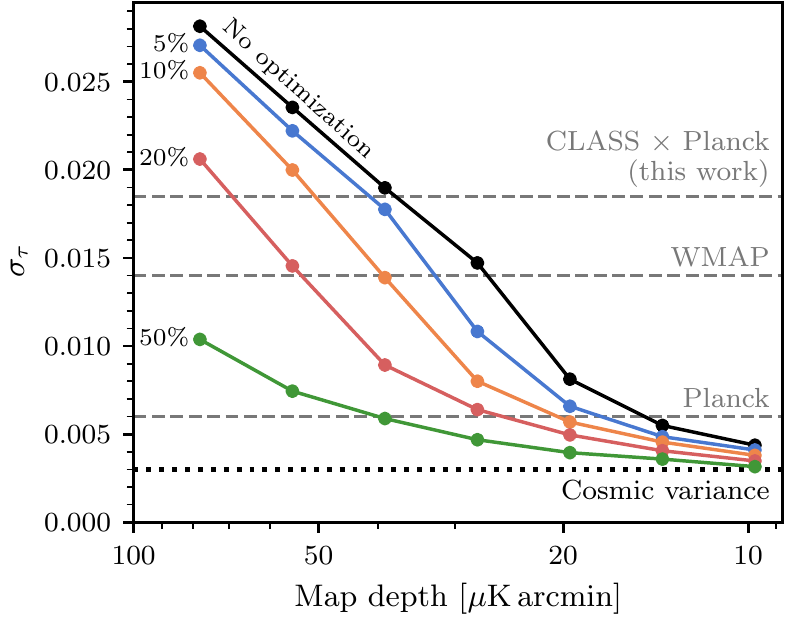}
        \caption{
            Forecast on the $\tau$ constraint precision, $\sigma_\tau$, from CLASS data alone as a function of improved map depth (decreasing map noise) assuming filtering optimizations.
            The black curve represents the forecast for the baseline analysis without filtering optimization; 
            each colored curve corresponds to a particular filtering optimization choice $\lambda$, with the $\lambda$ value labeled on the left side of the curve.
            The precision from CLASS $\times$ Planck (Section~\ref{ssec:tau-likelihood}), WMAP \citep{hinshaw13}, and Planck \citepalias[\sroll;][]{pagano19} are shown as horizontal dashed lines.
            The cosmic variance limit over approximately $40\%$ of the sky is shown as the dotted line.
        \label{fig:tau-forecast}}
\end{figure}

The $\sigma_\tau$ forecasts were derived from simulations following the procedure outlined in Section~\ref{ssec:tau-likelihood}. We generated 500 CMB-plus-noise simulations for two CLASS half-depth splits for each of the 600 parameter pairs. 
To focus solely on the performance improvement from filtering optimization and noise reduction, we ignored all foreground components and the uncertainties from foreground cleaning residuals.
Since two splits used for cross-correlation were both filtered, the xQML estimator differed from that used in Section~\ref{ssec:low-ell-spec} and applied transfer-matrix correction to both maps.
The $\tau$ posterior distributions were obtained for each of the 500 simulations that used the other 499 simulations to build the likelihood. The median $68\%$ intervals are reported as our projection results and shown in Figure~\ref{fig:tau-forecast}.

The map depth projections in Figure~\ref{fig:tau-forecast} are evaluated at $\sqrt{2}$ increments, corresponding to doubling the integration time at each step.  
Several near-term improvements are expected to achieve these gains in sensitivity, including recovery of half of the data discarded due to the closeout issue, a doubled linear polarization mapping speed via the RHWP \citep[due to not observing circular polarization,][]{chuss12-trans}, and the deployment of a second 90\ghz telescope.
Note that the current forecast considers only the cross-spectrum of two half-depth maps. In the noise-dominated regime, the sensitivity can be further improved by a factor of approximately two by partitioning the data into multiple splits---a strategy that becomes increasingly viable as sky coverage redundancy improves with future observations.
Figure~\ref{fig:tau-forecast} further emphasizes the importance of filtering optimization: less aggressive filtering would help surpass Planck constraints more quickly, and at least a 20\% enhancement is needed to approach the cosmic variance limit. 
That limit is around $0.003$ given the CLASS survey region and foreground masking, aligning with the results from \cite{watts18} and \cite{errard16}, despite differences in methodology.

\section{Conclusion}\label{sec:conclusion}
We presented the initial processing of the CLASS 90\ghz data collected from 2021 to 2024 with the VPM as the polarization modulator. 
The validity of both the data and the analysis pipeline has been verified through extensive null tests and by cross-correlating with external CMB maps.
With the exception of one null test failure---attributed to a noise model inadequacy under conditions where the instrument’s systematic cancellation was not fully leveraged---we found good consistency both internally and externally.

To address the signal bias due to time stream filtering, we carried out a detailed characterization of the pixel-space transfer matrix and described an updated implementation of the quadratic cross-spectrum estimator that works with the transfer matrix to correct for the filtering bias down to $\ell=2$.
Moreover, we investigated the non-linearity induced by the noise modeling in the maximum-likelihood map-maker and found a $\lesssim 3\%$ bias on the largest angular scale---this is the only bias that remains that was not fixed with linear correction.

Building upon this progress, we presented the first ground-based reionization optical depth measurement through a cross-correlation analysis with Planck 100 and 143\ghz maps.
The baseline result using multipole range $2\leq\ell\leq30$ with $50\%$ Galactic avoidance is $\tau = 0.053 ^{+0.018}_{-0.019}$, which is robust against choices of analysis mask, data set combination, and multipole selections. The null hypothesis of no reionization is rejected at $99.4\%$ confidence level.
The significance of $\tau$ measurements was reduced and only upper limits were reported when multipoles below $\ell=5$ were dropped, highlighting the importance of the largest angular scale recovery.

CLASS leads the ground-based effort of the largest-scale polarization measurement and has already demonstrated the effectiveness of polarization modulation and maximum-likelihood map-making in producing high-quality data. 
At the same time, CLASS’s pioneering systematic investigations---including this work---have revealed that time-domain filtering is necessary to address various forms of scan-correlated systematics.
Two major developments are essential to ensuring success: 1) a deeper understanding of the underlying systematics, ultimately enabling less aggressive mitigation, and 2) a reliable method for extracting cosmological signals in the presence of filtering. 
This work represents a step forward in the latter aspect, demonstrating that the transfer matrix can effectively correct linear biases in both foreground cleaning and power spectrum estimation.
Moreover, forecasts of CLASS’s ability to improve $\tau$ measurements further suggest that a modest 20\% optimization in the filtering strategy (see Section~\ref{sec:forecast}) could bring the experiment close to the cosmic variance limit.

Looking ahead, multiple efforts are underway toward these CLASS objectives. A detailed analysis of scan-correlated signals (\inprep{Chan et al.}) may guide refinements to the current filtering strategy and potentially help recover nearly half of the VPM data discarded in this analysis due to modeling challenges. 
Given the current aggressiveness of filtering, even incremental improvements could substantially enhance low-$\ell$ recovery.
In parallel, the CLASS 90\ghz telescope has been operating since June 2024 with an alternative modulator design \citep{eimer2022spie,shi24-spie}, expected to double the linear polarization mapping speed and address systematics associated with the VPM. 
Further, a second 90\ghz telescope is scheduled for deployment in 2025 to enhance CMB sensitivity substantially.

\section{Acknowledgments} The authors would like to thank Graeme Addison for helpful discussions and Luca Pagano and Reijo Keskitalo for sharing the Planck data products used for this work.
Y.L. is supported by the Kavli Institute for Cosmological Physics at the University of Chicago through an endowment from the Kavli Foundation.
R. D\"unner thanks ANID for grants BASAL CATA FB210003, FONDEF ID21I10236, and QUIMAL240004.
Z.X. was supported by the Gordon and Betty Moore Foundation through grant GBMF5215 to the Massachusetts Institute of Technology.
We acknowledge the National Science Foundation Division of Astronomical Sciences for their support of CLASS under grant Nos. 0959349, 1429236, 1636634, 1654494, 2034400, and 2109311.
We thank Johns Hopkins University President R. Daniels and Dean C. Celenza for their steadfast support of CLASS.  
We further acknowledge the very generous support of Jim Murren and Heather Miller (JHU A\&S '88), Matthew Polk (JHU A\&S Physics BS '71), David Nicholson, and Michael Bloomberg (JHU Engineering '64).
The CLASS project employs detector technology developed in collaboration between JHU and Goddard Space Flight Center under several previous and ongoing NASA grants. Detector development work at JHU was funded by NASA cooperative agreement 80NSSC19M0005. 

We acknowledge scientific and engineering contributions from Max Abitbol, Aamir Ali, Fletcher Boone, David Carcamo, 
Manwei Chan, Francisco Espinoza, 
Pedro Flux\'a Rojas, Joey Golec, 
Dominik Gothe, Ted Grunberg, 
Mark Halpern, Saianeesh Haridas, 
Gene Hilton, Connor Henley, 
Lindsay Lowry, 
Jeffrey~John McMahon, Nick Mehrle, 
Hayley Nofi,
Carolina N\'{u}\~{n}ez,
Keisuke Osumi,
Ivan~L. Padilla, Gonzalo Palma,  Bastian Pradenas, Isu Ravi, 
Carl~D. Reintsema, Gary Rhoades, Daniel Swartz, Bingjie Wang, Qinan Wang, Tiffany Wei, and Zi\'ang Yan. 
We thank Mar\'ia Jos\'e Amaral, Chantal Boisvert, William Deysher, Miguel Angel D\'iaz, Jill Hanson, Patrick Keating, and Joseph Zolenas for logistical support. 
We acknowledge the productive collaboration of the JHU Physical Sciences Machine Shop team. 

We acknowledge the use of the Legacy Archive for Microwave Background Data Analysis (LAMBDA), part of the High Energy Astrophysics Science Archive Center (HEASARC). HEASARC/LAMBDA is a service of the Astrophysics Science Division at the NASA Goddard Space Flight Center.
Part of this research project was conducted using computational resources of Advanced Research Computing at Hopkins (ARCH) and the National Energy Research Scientific Computing Center (NERSC). CLASS is located in the Parque Astron\'omico Atacama in northern Chile under the auspices of the Agencia Nacional de Investigaci\'on y Desarrollo (ANID).

\software{
numpy \citep{numpy20}, 
scipy \citep{scipy}, 
matplotlib \citep{matplotlib},
astropy \citep{astropy}, 
healpix \citep{healpix},
camb \citep{camb},
PolSpice \citep{polspice},
xQML \citep{xQML},
emcee \citep{emcee},
GPy \citep{gpy},
getdist \citep{getdist}
}

\bibliographystyle{aasjournal}
\bibliography{cosmology,references, software, class_common, cmb, foreground, hardware,Planck_bib}

\appendix
\section{Quadratic estimator with filtering correction}\label{app:xqml}

Following the notation in \cite{xQML}, the unnormalized quadratic estimator for the cross-spectrum of two maps $\mathbf{d}_A$ and $\mathbf{d}_B$ can be expressed as 
\begin{equation}
\hat{y}^{AB}_\ell = \mathbf{d}^T_A \mathbf{E}_\ell \mathbf{d}_B,
\end{equation}
where $\mathbf{E}_\ell$ is the estimator matrix for each multipole.
No noise bias subtraction is needed due to the absence of correlated noise between the maps.
For a given matrix $\mathbf{E}_\ell$, an unbiased estimate of the power spectrum can be recovered by normalizing the estimator as
\begin{equation}
    \hat{C}_\ell = \sum_{\ell^\prime} W^{-1}_{\ell\ell^\prime} \hat{y}^{AB}_\ell,
\end{equation}
where the normalization matrix is defined as
\begin{align}
    W_{\ell\ell^\prime} &= \text{Tr}\left[\mathbf{E_\ell}\frac{\partial \mathbf{S} }{\partial C_{\ell^\prime}}\right]\\
    &= \mathrm{Tr}[\mathbf{E_\ell}\mathbf{P}_{\ell^\prime}],
\end{align}
and $\mathbf{S}$ represents the signal covariance matrix common in both maps.
For a minimum-variance estimation of the spectrum, the estimator matrix is chosen to be 
\begin{equation}
    \mathbf{E}_\ell = \frac{1}{2}(\mathbf{C}^{AA})^{-1}\mathbf{P}_\ell(\mathbf{C}^{BB})^{-1}, \label{eq:xqml-el}
\end{equation}
where $\mathbf{C}^{AA}$ and $\mathbf{C}^{BB}$ are the total covariance in the two maps.

The original implementation of the estimator\footnote{\url{https://gitlab.in2p3.fr/xQML/xQML}} only supports isotropic filtering (e.g., the beam and pixel window functions) correction in harmonic space and common sky masking between the two maps.
To account for linear filtering that can be characterized by a pixel-space transfer matrix, we express the covariance of a filtered map as 
\begin{equation}
    \mathbf{C}^{AA} = \mathbf{M}_A\mathbf{F}_A\mathbf{S}\mathbf{F}_{A}^T\mathbf{M}^T_A + \mathbf{N}_A\label{eq:xqml-new-cov},
\end{equation}
where $\mathbf{F}_A$ is the pixel-space transfer matrix for map $A$.
A similar expression holds for map $B$, and the implementation supports both maps to be filtered and/or masked differently.
Hereon, we use the subscript $A$ for discussion because only map $A$ (the CLASS map) is filtered for the cross-spectrum estimation.
The rectangular masking matrix $\mathbf{M}_A$ of shape $(2 n_A, 2 n_\mathrm{F,A})$ describes the down-selection of pixels from $2n_\mathrm{F, A}$ in the filtered map to $2n_\mathrm{A}$ for cross spectrum estimation. 
The latter is also the number of pixels over which the noise covariance matrix $\mathbf{N}_A$ is defined.
The factor of 2 denotes the two linear polarization maps ($Q/U$) used for polarization analysis, but this can be easily generalized to include a temperature or circular polarization component.
Notably, Equation~\ref{eq:xqml-el} is the only place where the pixel-space covariance matrix is inverted. As such, even if the covariance matrix is ill-conditioned (e.g., due to missing modes from filtering), which is not the case here, partially because of the noise contribution, and requires pseudo-inversion, it would only impact the estimator’s optimality but would not introduce any bias.
With this modification, the two (inverse) covariance matrices in Equation~\ref{eq:xqml-el} can have different dimensions.
Therefore, the definition of $\mathbf{P}_\ell$ should be updated accordingly and be evaluated from a rectangular signal covariance matrix.
The normalization matrix is also updated to reflect the filtering correction to the signal covariance $\mathbf{S}^{BA}$:
\begin{align}
    W_{\ell\ell^\prime} &= \text{Tr}\left[\mathbf{E_\ell}\frac{\partial \mathbf{S}^{BA} }{\partial C_{\ell^\prime}}\right]\\
    &= \text{Tr}\left[\mathbf{E_\ell}
    \mathbf{M}_B\mathbf{F}_B\mathbf{P}_{\ell^\prime}\mathbf{F}_{A}^T\mathbf{M}^T_A
    \right].
\end{align}

\begin{figure}
    \includegraphics[width=\linewidth, ]{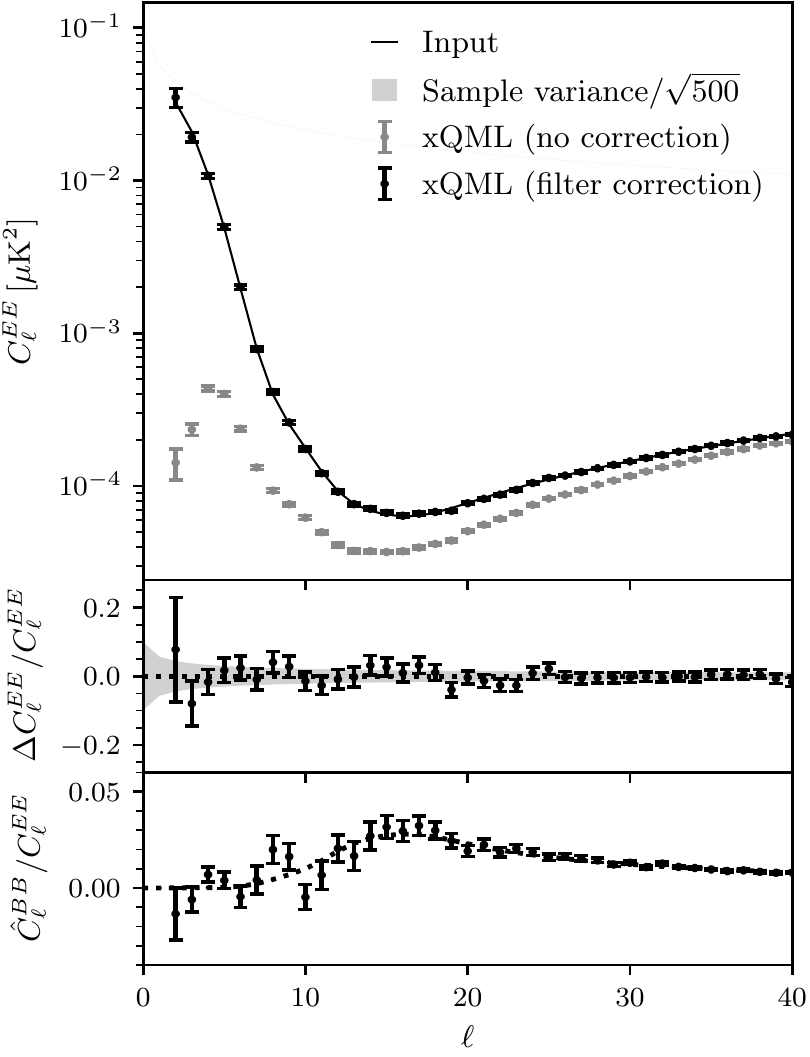}
    \caption{
    Validation of the filtering correction in the xQML implementation. 
    \emph{Top:} $EE$ power spectra of 500 signal-only fiducial $\Lambda$CDM reobservation simulations computed using xQML with (in black) and without (in gray) the filtering correction.
    The error bars reflect the uncertainty on the mean of the 500 simulations.
    For each simulation, the map is reobserved with the map-making pipeline, downgraded to $N_\mathrm{side}=16$, and cross-correlated with the simulation input directly downgraded to the same resolution.
\emph{Middle:} The fractional difference between the input and the filter-corrected $EE$ power spectra. For reference, the sample variance (of the mean of 500 simulations) is shown as gray shades. 
    The corrected power spectrum is unbiased up to $\ell=40$.
    \emph{Bottom:} The ratio between the filter-corrected $BB$ spectra and the input $EE$ spectra. The shape of the ratio is consistent with the input $B$-mode signal from CMB lensing (the dotted line), indicating minimal $E$-to-$B$ leakage. The variance in this case is much larger than the $B$-mode sample variance (not shown) or variance leakage from $E$ modes \citep{xQML} due to the suboptimal configuration of the xQML estimator for signal-only simulations.
    }
    \label{fig:xqml-validation}
\end{figure}

To validate this implementation, we tested the signal recovery using various combinations of masking and transfer matrices applied to either or both maps.
Figure~\ref{fig:xqml-validation} demonstrates an example that is a close analog to our baseline analysis: the filtered map (through the reobservation rather than the transfer matrix) is cross-correlated with an unbiased map, and the two maps are masked by the \texttt{f050} CLASS and Planck masks respectively (Figure~\ref{fig:masks}).
The same covariance matrices from Equation~\ref{eq:fg-red-N} (with appropriate filtering operator applied) are used for the covariance for xQML. 
While this is not the optimal choice for a signal-only spectrum, it ensures maximum comparability with the data.
With the masking operator introduced in Equation~\ref{eq:xqml-new-cov}, the transfer matrix can, in principle, be evaluated at higher $N_\mathrm{side}$ for a more accurate representation of the filtering effect, albeit with increased computational cost.
Here we opt to use the same $N_\mathrm{side}=16$ transfer matrix ($\mathbf{F}_{16,16}$), and the recovered $EE$ and $BB$ power spectra are unbiased up to $\ell=40$.
Since the filtered map is produced with the reobservation pipeline rather than the transfer matrix, the test here validates not only the xQML implementation itself but also checks the validity of using the pixel-space transfer matrix to approximate the time-domain filtering (reobservation).
Compared to the theoretical calculation of the sample variance (shaded region in the middle panel of Figure~\ref{fig:xqml-validation}), the inflated error bars from the simulation reflect filtering-induced mode loss and the sub-optimality of this particular xQML configuration (non-zero $\mathbf{N}_A/\mathbf{N}_B$) for signal-only spectra.

\section{VPM sync null test}\label{app:nulltest}
In this section, we provide more details on the single null test failure in the $EE$ spectrum of the VPM sync split.
As explained in previous studies on the VPM \citep{harrington2018thesis,harrington21}, the VPM emission (from both the wire grid and the mirror) and the unpolarized sky signal can be modulated at the mechanical frequency of the VPM due to non-idealities in the wire transmissions. This effect is commonly referred to as the VPM synchronous signal.
A key feature of the VPM is that linear polarization is modulated at the same frequency as the VPM motion. As a result, linear polarization demodulation can pick up this synchronous signal, leading to temperature-to-polarization (T-to-P) leakage. 
However, studies have shown that the VPM synchronous signal is stable over time, and its effect on demodulated data manifests primarily as a low-frequency drift. This drift occurs on time scales much longer than the crossing scale of the largest angular mode on the sky and is largely canceled out between orthogonal detector pairs.
Consequently, the VPM synchronous signal has been found to be a subdominant contributor to low-$\ell$ noise 
(\inprep{Cleary et al.,}), especially with moderate filtering and when the majority of detectors are paired.

\begin{figure}
    \centering
    \includegraphics[width=\linewidth]{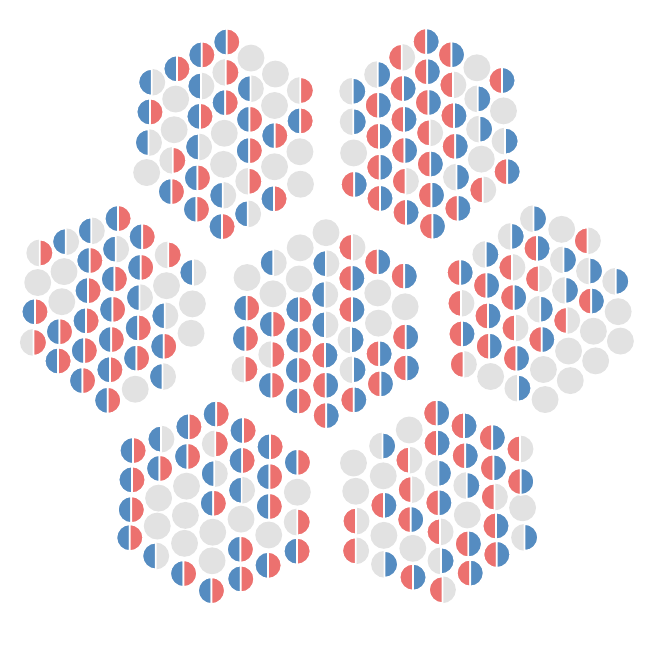}
        \caption{Detector assignment of the ``VPM sync'' split of the CLASS 90~GHz telescope.
        Each pair of semicircles represents an orthogonal pair of detectors, with the $+45^\circ$ and $-45^\circ$ orientations shown on the left and right, respectively.
        The detectors used for mapping in Era 3, featuring the upgraded focal plane, are color-coded according to their split assignment.
        The $+45^\circ$/$-45^\circ$ detectors on the left and right sides of the focal plane, shown in blue, are expected to observe higher amplitudes of VPM emission, while those in the other split, shown in red, are expected to observe lower amplitudes.
        \label{fig:vss-splits}}
\end{figure}

Nevertheless, the ``VPM sync'' detector split is designed to partially check for this. 
Given how each detector’s electric field (in a reciprocal sense) is projected onto the VPM wire, the $+45^\circ$-oriented detectors (relative to vertical, as viewed from the focal plane toward the sky) on the left of the focal plane and $-45^\circ$ detectors on the right side observe higher VPM synchronous signal absolute amplitudes than the other split.
This split is visualized in Figure~\ref{fig:vss-splits} for the Era-3 focal plane after the detector module upgrade. 
We note that this is purely a detector split based on the geometry rather than empirically determined from VPM signal measurement (though the measurement is qualitatively consistent with this model).

\begin{figure}[t]
    \centering
    \includegraphics[width=\linewidth]{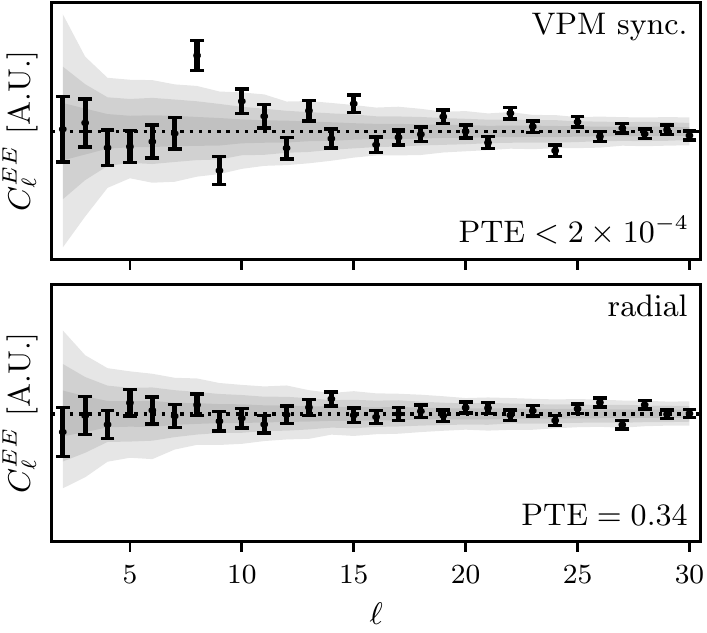}
        \caption{The low-$\ell$ $EE$ null spectra for the VPM sync split (top) and the radial split (bottom) for comparison.
        The shaded regions represent the 1/2/3-$\sigma$ range from noise-only simulations.
        \label{fig:null-spectra}}
\end{figure}

The low-$\ell$ $EE$ null spectrum for the VPM sync split is shown in Figure~\ref{fig:null-spectra}. 
This spectrum has the minimum PTE value below $2\times10^{-4}$, and the chance of seeing such an outlier among the total 114 spectra is less than $0.2\%$.
The low PTE is primarily pulled by the multipole $\ell=8$ ($5.1\sigma$), though several other multipoles in the $>2\sigma$ regime also contribute.
A possible explanation is that the noise model used to generate simulations is insufficient for this type of split, where systematics in single-detector data are not fully mitigated by pair-differencing.
For comparison, the bottom panel of Figure~\ref{fig:null-spectra} displays a better-behaved null spectrum for the ``radial'' detector split, which is based on the detector distance from the center of the focal plane. 
The noise spectra in these two cases reveal elevated noise power for $\ell < 20$ in the VPM sync split, which the simulation partially captures but fails to fully account for.

As detailed in \citetalias{Li23}, the noise model for demodulated data incorporates detector correlations through singular-value decomposition and models statistical properties using binned Fourier-space power spectra.
However, this approach may not adequately describe more complex (e.g., non-Gaussian) structures. 
These complexities can come from the VPM synchronous signal or other sources of T-to-P leakage-like effects, which can be largely canceled in pair-differencing.
However, the 90~GHz (Era 2) single-detector monopole T-to-P leakage is measured to be around $1.7\times10^{-3}$ with significant variation among modules \citep{datta23}. 
Consequently, residual leakage from unpolarized noise could remain in the VPM sync split (because the leakage from one side of the focal plane does not match exactly the other half), complicating the modeling.

In summary, the apparent low PTE value seen in the VPM sync split, particularly at $\ell=8$, indicates inadequate noise modeling when the instrument's systematic-canceling ability is compromised. This issue is unlikely to affect the final analysis, as most detectors in the final data are paired. Furthermore, the specific outlier at $\ell=8$ does not appear in the final CLASS--Planck cross-correlation analysis (Figure~\ref{fig:spec-xqml}), and the derived $\tau$ constraint is robust to the exclusion of this multipole (Figure~\ref{fig:tau}).

 \end{document}